\documentclass[sigplan, screen, nonacm]{acmart}
\usepackage{enumitem}
\usepackage[algo2e,ruled,lined,linesnumbered,noend]{algorithm2e}
\usepackage{url}
\usepackage{listings}
\usepackage{cleveref}
\usepackage{soul}

\ifx\nocomment\undefined
\newcommand{\guy}[1]{{\color{brown}{\bf Guy:} #1}}
\newcommand{\yan}[1]{{\color{violet}{\bf Yan:} #1}}
\newcommand{\yihan}[1]{{\color{blue}{\bf Yihan:} #1}}
\newcommand{\laxman}[1]{{\color{purple}{\bf Laxman:} #1}}
\newcommand{\magdalen}[1]{{\color{magenta}{\bf Magdalen:} #1}}
\newcommand{\zheqi}[1]{{\color{cyan}{\bf Zheqi:} #1}}
\else
\newcommand{\guy}[1]{{}}
\newcommand{\yan}[1]{{}}
\newcommand{\yihan}[1]{{}}
\newcommand{\laxman}[1]{{}}
\newcommand{\magdalen}[1]{{}}
\newcommand{\zheqi}[1]{{}}
\fi

\newcommand{\hide}[1]{}

\newcommand{\algoname}[1]{\textsf{#1}}

\newcommand{\sysname}{\algoname{ParlayANN}}
\newcommand{\diskann}{\algoname{DiskANN}}
\newcommand{\hnsw}{\algoname{HNSW}}
\newcommand{\hcnng}{\algoname{HCNNG}}
\newcommand{\pynn}{\algoname{PyNNDescent}}
\newcommand{\pardiskann}{\algoname{ParlayDiskANN}}
\newcommand{\parhnsw}{\algoname{ParlayHNSW}}
\newcommand{\parhcnng}{\algoname{ParlayHCNNG}}
\newcommand{\parpynn}{\algoname{ParlayPyNN}}
\newcommand{\faiss}{\algoname{FAISS}}
\newcommand{\falconn}{\algoname{FALCONN}}

\newcommand{\BIGANN}{\text{BIGANN}}
\newcommand{\SPACEV}{\text{MSSPACEV}}
\newcommand{\TTI}{\text{TEXT2IMAGE}}
\newcommand{\SSNPP}{\text{SSNPP}}
\newcommand{\Nout}{N_{\text{out}}}
\newcommand{\mL}{\mathcal{L}}
\newcommand{\mV}{\mathcal{V}}
\newcommand{\mP}{\mathcal{P}}
\newcommand{\mK}{\mathcal{K}}
\newcommand{\mQ}{\mathcal{Q}}
\newcommand{\la}{\leftarrow}

\providecommand{\norm}[1]{\lVert#1\rVert}
\newcommand{\dist}[2]{{\norm{#1,#2}}}

\newcommand{\batchupperbound}{\theta}

\newcommand{\defn}[1]{{\emph{\textbf{#1}}}}
\newcommand{\emp}[1]{\emph{\textbf{#1}}} 
\newcommand{\myparagraph}[1]{\vspace{0.2em}\noindent\emp{#1.} \,}
\newcommand{\codeskip}{\vspace{.3em}}
\newcommand{\mathtext}[1]{\mathit{#1}}

\SetCommentSty{mycommfont}

\makeatletter
\patchcmd{\@algocf@start}
  {-1.5em}
  {0pt}
  {}{}
\usepackage{cleveref}
\crefname{section}{Sec.}{Sec.}
\crefname{theorem}{Thm.}{Thm.}
\crefname{lemma}{Lem.}{Lem.}
\crefname{corollary}{Col.}{Col.}
\crefname{table}{Tab.}{Tab.}
\crefname{algorithm}{Alg.}{Alg.}
\crefname{figure}{Fig.}{Fig.}
\crefname{fact}{Fact}{Fact}
\Crefname{table}{Tab.}{Tab.}
\crefname{problem}{Problem}{Problem}

\acmYear{2024}\copyrightyear{2024}
\setcopyright{rightsretained}
\acmConference[PPoPP '24]{The 29th ACM SIGPLAN Annual Symposium on Principles and Practice of Parallel Programming}{March 2--6, 2024}{Edinburgh, United Kingdom}
\acmBooktitle{The 29th ACM SIGPLAN Annual Symposium on Principles and Practice of Parallel Programming (PPoPP '24), March 2--6, 2024, Edinburgh, United Kingdom}
\acmDOI{10.1145/3627535.3638475}
\acmISBN{979-8-4007-0435-2/24/03}

\usepackage{booktabs}   
\usepackage{subcaption} 

\setcopyright{none}
\renewcommand\footnotetextcopyrightpermission[1]{} 

\usepackage{float}
\usepackage[labelfont=bf,font={small},aboveskip=0em, belowskip=0em]{caption}

\usepackage{titlesec}

\titlespacing{\section}{0pt}{0.3em}{0.2em} 
\titlespacing{\subsection}{0pt}{0.3em}{0.15em} 
\titlespacing{\subsubsection}{0pt}{0.3em}{0.5em} 

\begin{document}

\title[ParlayANN: Parallel Graph-Based Approximate Nearest Neighbor Search Algorithms]{ParlayANN: Scalable and Deterministic Parallel Graph-Based Approximate Nearest Neighbor Search Algorithms}         



\author{Magdalen Dobson Manohar}       
\affiliation{
  \institution{Carnegie Mellon University} 
  \country{}           
}
\email{mrdobson@cs.cmu.edu}          

\author{Zheqi Shen}       
\affiliation{
	\institution{UC Riverside}
	\country{}             
}
\email{zshen055@ucr.edu}          

\author{Guy E. Blelloch}       
\affiliation{
	\institution{Carnegie Mellon University}   
	\country{}          
}
\email{guyb@cs.cmu.edu}          

\author{Laxman Dhulipala}       
\affiliation{
	\institution{University of Maryland} 
	\country{}            
}
\email{laxman@umd.edu}          

\author{Yan Gu}       
\affiliation{
	\institution{UC Riverside}
	\country{}             
}
\email{ygu@cs.ucr.edu}          

\author{Harsha Vardhan Simhadri}      
\affiliation{
	\institution{Microsoft Research}  
	\country{}           
}
\email{harshasi@microsoft.com}          

\author{Yihan Sun}       
\affiliation{
	\institution{UC Riverside}  
	\country{}           
}
\email{yihans@cs.ucr.edu}          

\renewcommand{\shortauthors}{Manohar, Shen, Blelloch, Dhulipala, Gu, Simhadri, and Sun}

\begin{abstract}
Approximate nearest-neighbor search (ANNS) algorithms are a key part
of the modern deep learning stack due to enabling efficient similarity
search over high-dimensional vector space representations (i.e.,
embeddings) of data.
%
%
Among various ANNS algorithms, graph-based algorithms are known to
achieve the best throughput-recall tradeoffs.
Despite the large scale of modern ANNS datasets, existing parallel
graph-based implementations suffer from significant challenges to scale to
large datasets due to heavy use of locks and other sequential bottlenecks, 
which 1) prevents them
from efficiently scaling to a large number of processors, and 2)
results in non-determinism that is undesirable in certain applications.

In this paper, we introduce ParlayANN, a library of deterministic and parallel graph-based approximate nearest neighbor search algorithms, along with a set of useful tools for developing such algorithms. In this library, we develop novel parallel implementations for four
state-of-the-art graph-based ANNS algorithms that scale to
billion-scale datasets.
Our algorithms are deterministic and achieve high scalability across
a diverse set of challenging datasets.
In addition to the new algorithmic ideas, we also conduct a detailed
experimental study of our new algorithms as well as two existing
non-graph approaches. Our experimental results both validate the
effectiveness of our new techniques, and lead to a comprehensive
comparison among ANNS algorithms on large scale datasets with a list
of interesting findings.

\end{abstract}

\ccsdesc[500]{Computing methodologies~Shared memory algorithms}
\ccsdesc[500]{Information systems~Retrieval tasks and goals}

\keywords{nearest neighbor search, vector search, parallel algorithms}  

\maketitle

\section{Introduction}\label{sec: intro}
The adoption of deep learning methods over the past decade have led to
high-dimensional vector representations of objects a.k.a.\ embeddings
becoming widely used. These representations are typically obtained by
training deep neural networks. As a result, machine learning datasets
usually contain billions of vectors representing embeddings of users,
documents, search queries, images, among many other kinds of objects. These
embeddings can span hundreds to thousands of dimensions.
The algorithms producing these embeddings are trained so that similar
objects have ``close'' embeddings (e.g., in $L_2$ distance). As a
result, an important problem is to find the nearest and thus most
similar set of $k$ objects for a query point in the embedding space
$\mathbb{R}^{d}$.

This problem is known as {\em $k$-nearest neighbor search},
and is notoriously hard to solve exactly in high-dimensional
spaces~\cite{beygelzimer2006cover}.
Since solutions for most real-world applications can tolerate small
errors, most deployments focus on the {\em approximate nearest
neighbor search (ANNS)} problem, which has been widely applied as a
core subroutine for search recommendations, machine learning, and
information retrieval~\cite{wang2021comprehensive}, as well as large
language models (LLMs) used in ChatGPT~\cite{chatgpt} and other
applications combining LLMs and vector search~\cite{chase2023vector,
tawalke2023pr, bing, stallbaumer2023copilot}.
Considering that embeddings and similarity search are at the heart of
these and many other modern AI applications, it is increasingly
important to build scalable and efficient parallel ANNS solutions that
can scale to massive modern datasets.

Some of the best-performing ANNS algorithms today are {\em
graph-based} ANNS algorithms, which are able to achieve high
recall (i.e., fraction of the true $k$-NNs returned by the query)
while obtaining high throughput (queries per second, or QPS).
Graph-based ANNS algorithms construct a {\em proximity graph} over the
points that connects each point with closeby points.  ANNS queries
search for the $k$-nearest neighbors of a query point by traversing
the proximity graph from a seed point, greedily exploring points that
are closer to the query until the search converges.
\hide{Graph-based algorithms outperform earlier approaches based on
clustering and locality-sensitive hashing~\cite{amuller2020ann,
wang2021comprehensive}\zheqi{LSH is one of the clustering-based methods} in practice across a wide range of datasets and
achieve superior recall and QPS compared to non-graph approaches, as
shown in many recent benchmarking
papers~\cite{subramanya2019diskann,munoz2019hcnng,mcinnes2020pynndescent,fu2019nsg,malkov2020hnsw}.
}
Among various types of ANNS algorithms, graph-based algorithms in general achieve superior recall and QPS,
as shown in many recent benchmarking papers~\cite{subramanya2019diskann,munoz2019hcnng,mcinnes2020pynndescent,fu2019nsg,malkov2020hnsw}.

\begin{figure}
	\centering
	\includegraphics[width=\columnwidth]{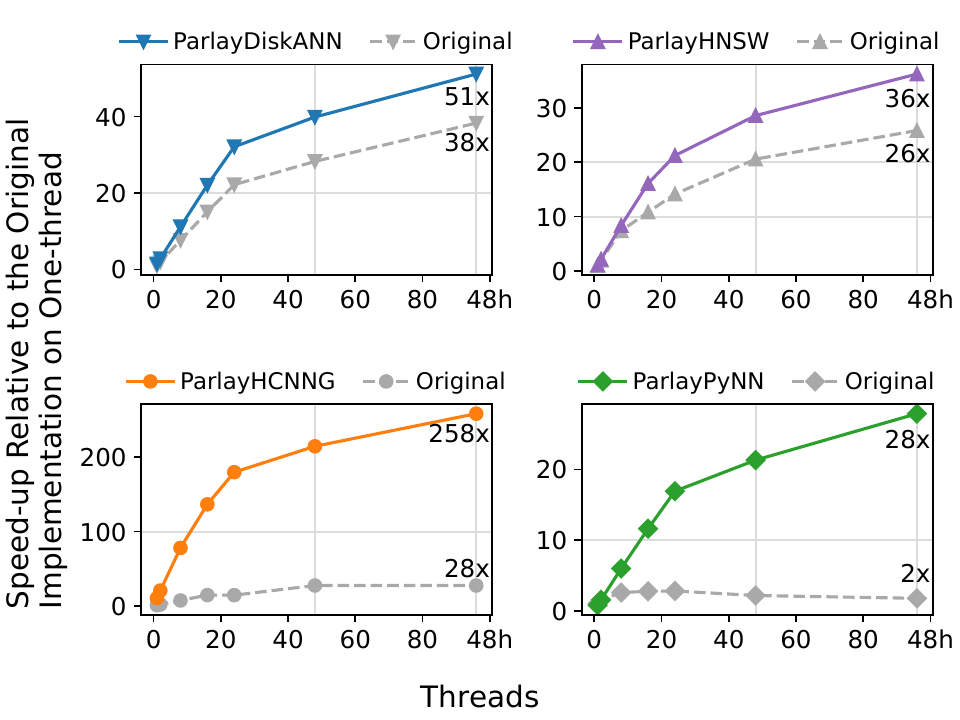}
  \caption{\small  {\bf Scalability of original and our new
  implementations of four ANNS algorithms on various number of
  threads.  Within each subfigure, all numbers are
  {\textit{speedup numbers relative to the original implementation on one
  thread}}. Higher is better.} 
Results were tested on a machine with 48 cores using dataset \BIGANN-1M ($10^6$ points). ``48h'': 48 cores with hyperthreads.
The two implementations in the same subfigure always use the same parameters and give similar query quality (recall-QPS curve).
\label{fig:buildparallelism}
	}
\end{figure}

Despite the focus on efficiency and benchmarking in the ANNS
literature, {\em there is very little work (algorithmic ideas or
benchmarking) that systematically studies how parallel
graph-based ANNS algorithms perform as we scale the input size and the
number of processors.}
On the algorithmic side, some graph-based algorithms do have parallel
implementations, but rely on per-vertex locks to enable parallelism
which raises two major issues affecting both performance and
``correctness''.
First, due to the use of locks, most existing implementations {\em tend to only scale
well to tens of threads}.
\cref{fig:buildparallelism} demonstrates parallel scalability curves for four state-of-the-art
(SOTA) implementations of graph-based algorithms (grey lines), on a well-known ANNS benchmark~\cite{amuller2020ann} with 1M points.
%
None of them achieve significant speedup beyond 50 threads.
Furthermore, using locks results in {\em non-deterministic} outputs,
i.e., multiple runs of the algorithm may
yield different proximity graphs due to lock acquisition order.
%
Non-determinism can be a serious issue for applications that require
persistence, crash recovery, or replication, e.g., for vector
databases such as Pinecone, Weaviate, and Lucene~\cite{pinecone,
weaviate, lucene}.


On the benchmarking side, existing
results~\cite{amuller2020ann,wang2021comprehensive} focus on
relatively small input sizes (usually
million-scale), and evaluate algorithms based on their
sequential performance.
%
Therefore, techniques that perform well on existing ANNS benchmarks may not be
suitable (or are unclear to be suitable) for a significantly larger
dataset or more cores.
Due to the lack of benchmarking studies focusing on parallelism,
we also find that some of the scalability issues for existing parallel implementations
are from some sequential bottlenecks that do not appear
until a large number of cores or sockets are used, or until they are
run on much larger datasets.
%
Therefore, understanding how different ANNS algorithms
scale from
million to billion-scale as a function of the number of cores, and
across a diverse set of datasets is an important open problem.



\textbf{To address this problem, in this paper we develop
\sysname{}, a parallel ANNS library
that scales to \emph{billion-scale} datasets, scales to more than a hundred
threads, and is deterministic.}
To achieve these goals, 
we exploit multi-threading (specifically, using
fork-join parallelism) as much as possible to
reduce the build time, which can be weeks on a single thread at such a scale.
%
We provide new general techniques for building ANNS graphs in
parallel, such as prefix doubling and batch updates.
We then apply our general techniques
to four SOTA graph-based algorithms:
\diskann{}~\cite{subramanya2019diskann},
\hnsw{}~\cite{malkov2020hnsw}, \hcnng{}~\cite{munoz2019hcnng} and
\pynn{}~\cite{mcinnes2020pynndescent}.
In addition to
new general techniques, we also developed several algorithmic
optimizations to remove scalability bottlenecks for each specific
algorithm, such as very large per-thread hash-tables (in \hnsw{}, see~\cref{sec:hnsw}),
and certain data structures overflowing the
L3 cache (in \hcnng{}, see~\cref{sec:hcnng}).
Our implementations, \pardiskann, \parhnsw{}, \parhcnng{} and
\parpynn{}, are {\em deterministic, and achieve much better
scalability than the best existing parallel implementations for each
of them}.

Many of the tools in our library are of general use; to give an idea of the generality and practicality of ParlayANN, ParlayANN contains about 5000 lines of code, of which around 2000 are specific to one algorithm and the remaining 3000 are shared. 

In Figure~\ref{fig:buildparallelism}, we present the scalability
of our implementations relative to existing implementations
of graph-based ANNS algorithms on 1M points (all numbers are relative
to the one-thread running time of the original implementation in each
kind).
Our implementations scale well up to all 48 cores on the machine we use, with
further performance improvements from hyperthreading.




We carefully benchmarked our new implementations along with two
existing SOTA non-graph algorithms
(\faiss{} and \falconn~\cite{andoni2015practical, douze2011product})
on diverse real-world datasets with a billion
points, including one dataset for
out-of-distribution (OOD) queries (see more details in~\cref{sec: experimental}).
Three of our implementations (\pardiskann{}, \parhnsw{} and
\parhcnng{}) scale to billion-size datasets with reasonable
preprocessing time for index building (around 10h) with high-quality
query results (up to .99 recall with about $10^4$ QPS).
Our graph-based implementations achieve the best tradeoffs between
recall and QPS across the recall spectrum, while the
non-graph approaches failed to achieve a recall higher than 95\% on
billion-size datasets, even with very low QPS.
%
%
{\bf We believe this is the first work that scales deterministic
parallel ANNS algorithms to billions of points with high recall.}

By supporting these algorithms in a unified framework (e.g., same
parallel framework, primitives, and work-stealing scheduler) and
applying similar optimization effort across all of them, our results
also provide a {\em fair comparison of the algorithmic ideas} among
the existing graph-based approaches, both for index quality and their
potential for parallelism.
Benchmarking these algorithms at a billion-scale required significant
programmer and computational effort; for example, building all of the
ANNS indexes shown in \cref{sec: experimental} (six algorithms
each with three datasets) took more than 90 hours of computation time on a machine with 64 cores.
Our efforts led to a variety of interesting new findings about how
ANNS algorithms perform as dataset sizes are scaled.
{\bf We believe this work is also the first to depict an accurate
picture of performance comparison among ANNS algorithms on
billion-scale datasets. }

In summary, our results include both algorithmic contributions and new
experimental findings about the performance of ANNS algorithms at very
large scales, listed as follows. We plan to release our code.
Due to page limits, we provide the full paper with appendix in the supplemental material.

\begin{enumerate}[label=\arabic*.,topsep=0pt,itemsep=0pt,parsep=0pt,leftmargin=15pt]
\item A variety of general and specific techniques to parallelize existing graph-based ANNS
algorithms to scale to billions of points (\cref{sec:algsurveyed}).

\item High-performance parallel implementation \sysname{},
which contains four graph-based ANNS algorithms. 

\item In-depth experimental study of existing and our algorithms on a variety of billion-scale datasets, including
a special dataset for out-of-distribution queries (\cref{sec: results}).

\item A list of interesting findings about parallel ANNS algorithms on large scale datasets (\cref{sec: results}).

\hide{\item {\bf Out-of-distribution datasets.}
For out-of-distribution datasets, i.e., when the query
distribution is from an inherently different source from the base
distribution (e.g., text queries on images-embedded datasets), we find
that {\em graph-based methods significantly outperform IVF and
LSH-based methods on billion-scale datasets}, which struggle with out-of-distribution data.

\item {\bf Range-query datasets.}
ANNS applications such as plagiarism or misinformation
detection~\cite{simsearchnet} require returning candidates within a
certain {\em range} of a query point instead of the $k$ closest
candidates. Even though standard ANNS  algorithms are easily adapted
to serve range queries, ANNS algorithms are not frequently evaluated
on range search datasets. In our work, we found that {\em IVF
methods have a significant advantage on range search datasets}.}

\end{enumerate}

\section{Preliminaries}
\label{sec:prelims}




\myparagraph{Parallel Model}
We use the fork-join model of parallelism~\cite{CLRS,blelloch2019optimal}.
We assume a set of threads that access a shared memory.
A process can \textsf{fork} two child software threads to work in parallel.
When both children complete, the parent process continues.
A parallel for-loop over $n$ items can be simulated by recursively
\textsf{forking} $\log n$ levels.
Computations in the model can be efficiently executed using a
randomized work-stealing scheduler~\cite{blumofe1999scheduling,ABP01}.

We say a parallel computation is \emp{deterministic} if it gives the
same output across multiple runs, i.e., the output is not affected by
the runtime scheduler.
For randomized algorithms, we assume the randomness is supplied as
part of the input (e.g., as a random seed).

\myparagraph{Parallel Semisort} Many of our algorithms use a parallel
semisort~\cite{gu2015top} as a subroutine.  Given a sequence $A$ of entries, each
associated with a \emph{key}, a semisort reorders $A$ such that all
entries with the same key are consecutive. Note that the entries or
keys do not need to be fully sorted.

\myparagraph{Approximate Nearest Neighbor Search (ANNS)} In this work,
we study a set $\mP\subseteq \mathbb{R}^{d}$ of $n$ points (vectors)
in $d$ dimensions.
We denote the \defn{distance} between two points $p,q\in
\mathbb{R}^{d}$ as $\dist{p}{q}$.  Smaller distance indicates greater
similarity.
Commonly-used distance functions include Euclidean distance ($L_2$
norm), and cosine distance $(1 - \cos(\theta))$.

%
%

\begin{definition}($k$-NNS)
Given a set of points $\mP$ in $d$-dimensions and a query point $q$,
the $k$ nearest neighbor search ($k$-NNS) problem finds a set $\mK\subseteq \mP$ with size $|\mK|=k$,
such that $\max_{p\in \mK} \dist{p}{q} \le \min_{p\in \mP\setminus\mK} \dist{p}{q}$.
\end{definition}

We define $k$-ANNS as $k$-approximate NNS. With clear context, we omit $k$ and call them NNS and ANNS. We now introduce the most commonly-used measure of accuracy for ANNS, frequently referred to as \textit{recall}.

\begin{definition}($k@k'$ recall)
Let $\mP$ be a set of points in $d$-dimensions and $q$ a query point. Let $\mK$ be the true $k$-nearest neighbors of $q$ in $\mP$.
Let $\mathcal{\mK'} \subset \mP$ be an output of an ANNS algorithm of size $k'$. Then the $k@k'$ recall of $q$ is $\frac{|\mK \cap \mK'|}{|\mK|}$.
\end{definition}

The most common choice of recall is 10@10 recall.
Throughout the paper, 
we use the term ``recall'' to refer to 10@10 recall of an entire query set, i.e., the average recall over all points in the query set.

\section{General Techniques for Graph-Based ANNS Algorithms}\label{sec:algsurveyed}\label{sec:general}

In this section, we describe our new techniques for parallel graph-based algorithms.
We first present the high-level idea underpinning graph-based
ANNS algorithms.
We then introduce two major existing approaches: \emph{incremental
algorithms} and \emph{clustering trees},
as well as our new general techniques to make them parallel and deterministic.
%
In the next section, we show how these general techniques can be applied to
four graph-based ANNS algorithms.

\begin{figure}
  \centering
  \includegraphics[width=\columnwidth]{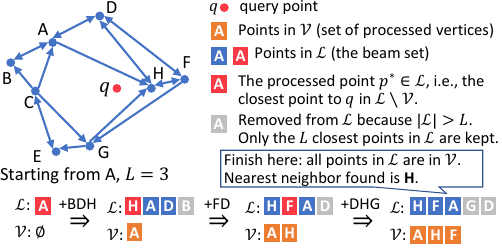}
  \caption{\small An example of ANNS graph and a greedy search. The blue arrows represent directed edges in the proximity graph, which is a mix
  of long and short edges. Below is an example of NNS query on point $q$ (red point).
  The algorithm starts with adding the starting point $A$ as the only point in the beam $\mathcal{L}$, and then in every step,
  finds the closest unprocessed point in $\mathcal{L}$ (to $q$)
  and adds its out-neighbors. Once $|\mathcal{L}|$ goes beyond $L$, it is refined to keep only the $L$ nearest points.
  A set $\mathcal{V}$ is maintained for all processed vertices.
  When all vertices in $\mathcal{L}$ are also in $\mathcal{V}$, the algorithm finishes.
  }\label{fig:example}
\end{figure}

\myparagraph{High-Level Approach}
Given point set $\mP$, an \emp{ANNS graph} $G_{\mP}$ refers to a
directed graph with vertices representing points in $\mP$.
For a point $p \in \mP$, we define $\Nout(p)$, or the out-neighbors of $p$.
We illustrate an example of an ANNS graph on them in \cref{fig:example}.
The neighborhood of a point in the graph roughly corresponds to other
nearby points, while some ``long edges'' are also needed (see details
below).

\myparagraph{Greedy (Beam) Search}
Almost all ANNS graph algorithms use a variant of \emp{greedy (beam)
search} to answer NNS queries (see \cref{fig:example} and \cref{algo:
greedySearch}).
Such a search for a query $q$ maintains a \emph{beam} $\mathcal{L}$
with size at most $L$ as a set of nearest neighbor candidates of $q$.
We call $L$ the \emph{width} of the beam.
The beam starts with a single starting point~$s$.
In each step, the algorithm pops the closest vertex to $q$ from
$\mathcal{L}$, and \emph{processes} it by adding all its out-neighbors
to the beam.
We use a \emph{visited set} $\mathcal{V}$ to maintain all points that
have been processed (i.e., the neighborhood of the point has been
traversed and added to the beam).
If $|\mathcal{L}|$ exceeds $L$, the $L$ closest points are kept.
%

Intuitively, for greedy search to converge quickly and produce
accurate answers, the ANNS graph should contain a mix of long edges
(connecting with neighbors that are far away) and short edges
(connecting with neighbors that are close).  Long edges enable fast
navigation from the starting point towards the region close to a query
point, and short edges enable the search to quickly converge once it
reaches this region of the graph.

\subsection{{Incremental Algorithms}}\label{sec:general:incremental}
One class of graph-based ANNS algorithms is \emph{incremental}
algorithms, which work by inserting all points into the graph in some
order; when inserting $p$, the algorithm adds new edges between $p$
and the existing points in the graph so that $p$ can be
discovered by queries.
Among the algorithms we study, \diskann{} and \hnsw{} are incremental
algorithms.
%

Most incremental graph algorithms, such as \diskann{}, \hnsw{}, and
\algoname{NSG}~\cite{subramanya2019diskann, malkov2020hnsw, fu2019nsg}
use a greedy search procedure as a substep during insertion.
\cref{algo: insert} presents the high-level idea of this \texttt{insert} routine.
Inserting a point $p$  (\cref{algo: insert}) first does a greedy search on the existing graph, and then
chooses the out-neighbors of $p$ from the visited set $\mV$ of the
search by performing a \texttt{prune} routine.
The \texttt{prune$(p,\mV,R)$} routine selects a subset from a
candidate set $\mV$ as the neighbors of $p$, which ideally should
cover a diverse range of edge lengths and directions.
Pruning also ensures that the size of $\Nout(p)$ has at most a given
\emp{degree bound} $R$; smaller $R$ typically results in fast but less
accurate searches compared to a larger $R$.
In addition to selecting out-neighbors of $p$, the \texttt{insert}
algorithm must add $p$ as the out-neighbors of other points so $p$ is
reachable during a search.
This is done by adding $p$ to each of $p$'s out-neighbor $q$, and calling
\texttt{prune} on $q$ to ensure the degree bound $R$.
The pruning strategies are specific to each graph-based algorithms, and we
describe them in \cref{sec:algos}.

\myparagraph{{Challenges for Incremental Algorithms}} To parallelize
incremental ANN algorithms, many existing implementations (e.g.,
\diskann{}) insert all points in a single parallel loop over
\emph{all} the points, with per-point locks
to ensure that the points are accessed safely.
This can cause 
performance issues and cause non-determinism.
%

\begin{algorithm2e}[t]
	\caption{greedySearch($p, s, L, k$).\protect}
	\label{algo: greedySearch}\small
	\SetKwBlock{ParDo}{do in parallel}{end}
	\SetKwBlock{ParFor}{parallel for}{end}
	\SetAlgoLined
    \DontPrintSemicolon
	\KwIn{Point $q$, starting point $s$, beam width $L$, integer $k$.}
	\KwOut{Set $\mathcal{V}$ of visited points and set $\mathcal{K}$ of $k$-nearest neighbors to point $q$. }

	$\mV \; \la \; \emptyset$ \;
	$\mL \; \la \; \{s\}$ \;
	\While{$\mL\setminus \mV \not= \emptyset$}{
		$p^* \; \la \; \arg\min_{(p \in \mL\setminus \mV)} \norm{p, q}$\\
		$\mV \; \la \mV \cup \{p^*\}$  \;
		$\mL \; \la \; \mL \cup \Nout(p^*)$\label{line:addneighbor}\\
		\lIf{$|\mL| > L$}{
			retain only $L$ closest points to $q$ in $\mL$
		}
	}
	$\mathcal{K} \; \la \; k$ closest points to $q$ in $\mV$\;
	\KwRet{$\mV, \mathcal{K}$}\;
\end{algorithm2e}

\begin{algorithm2e}[t]
	\caption{insert($p, s, R, L$).\protect}
	\label{algo: insert}\small
	\SetKwBlock{ParDo}{do in parallel}{end}
	\SetKwBlock{ParFor}{parallel for}{end}
    \DontPrintSemicolon
	\SetAlgoLined
	\KwIn{Point $p$, starting point $s$, beam width $L$, degree bound $R$.}
	\KwOut{Point $p$ is inserted into the nearest neighbor graph.}
	$\mV, \mathcal{K} \; \la \; \text{greedySearch}(p, s, L, 1)$  \;

	$\Nout(p) \; \la \; \text{prune}(p,\mV,R)$  \;
	\For{$q \in \Nout(p)$}{
		$\Nout(q) \; \la \; \Nout(q) \cup \{p\}$ \;
		\lIf{$|\Nout(q)| > R$}{
			$\Nout(q) \; \la \; \text{prune}(q,\Nout(q),R)$
		}
	}
\end{algorithm2e}

\myparagraph{New Technique in ANNS: Prefix Doubling}
\hide{As mentioned, many existing parallel implementations (e.g., \diskann{}
and \hnsw{}) of incremental ANNS algorithms run insertions of
\emph{all} points in a parallel for loop  with per-point locks used to
ensure that the out-neighbors of each point are accessed safely.}
%
We now present our first technique to avoid using locks in incremental graph-based algorithms.
Note that the main reason of using locks in the existing implementations is that the points being inserted in parallel all start from an \emph{empty} index (graph),
and therefore need a way to ``see'' each other and to ``bootstrap''.
Using locks effectively sequentializes all conflicts and achieve a result close to the sequential algorithm,
but introduces performance and non-determinism issues.

To address this, we use \emph{prefix-doubling}~\cite{blelloch2016parallelism,gu2023parallel,gu2022parallel,shen2022many,gbbs2021}.
%
The high-level idea is to insert points in batches of exponentially
increasing size (but upper bounded by a parameter $\batchupperbound$,
see details below), as shown in the while-loop in \cref{algo:
batchBuild}.
Each point will add itself based on the \emph{snapshot at the end of
the last batch}, and therefore points do not conflict with each other.
Initially, the batches are relatively small, which more closely
resembles the sequential version, allowing for a more accurate index
initially.
When the index becomes reasonably large, larger batches are allowed,
which also enables high parallelism.
Compared to the sequential version where point $i$ is inserted based
on the index of $i-1$ points, this approach allows point $i$ to
deterministically see an index with $O(i)$ points (roughly $i/2$),
while extracting significant parallelism.
For potential conflicts when adding multiple points to the
neighborhood of an existing point, we carefully merge them together
using a deterministic semisort.
Prefix-doubling provides balance between \emph{parallelism} (most of the
batches are sufficiently large to utilize a large number of threads),
\emph{progress} (no contention or race within each batch), and
\emph{accuracy} (each point see a reasonably accurate snapshot of the
index).

\myparagraph{New Technique in ANNS: Batch Insertion and Pruning}
A basic building block in our incremental algorithms is \emp{batch
insertion}, which adds a batch of points to the current index.
In \cref{algo: batchBuild}, inserting each batch involves two steps:
(1) building the neighborhood for the newly-inserted points (Lines
\ref{line:build:for1:start}--\ref{line:build:for1:end}), and (2)
adding the reversed edges to the existing points (Lines \ref{line:build:for2:start}--\ref{line:build:for2:end}).
%
Step 1 deals with each point in the batch in parallel, which uses a
greedy search on the immutable snapshot index to find a candidate set,
followed by pruning the candidates.
In this step, all points in the batch construct their own neighborhood independently on an immutable
snapshot, and thus does not affect each other.
Therefore, this step is parallel and deterministic, and no locks are needed.
\hide{\laxman{Parallelism by virtue of the prefix-doubling? Maybe we don't
need to mention parallelism here since we cover it while talking about
prefix doubling.}
}

In the next step, the edges are reversed and any vertices whose
neighborhood exceeds the degree threshold are pruned.
To do this in deterministic manner without using locks, we collect all
edges to be added in $\mathcal{B}$ in the format $(u,v)$, where $u$ is
a newly-added point in this batch, and $v$ is an existing point in the
graph.
We then run a \emp{parallel semisort} (see \cref{sec:prelims}) on
$\mathcal{B}$ by the key of $v$, such that all edges incident the same
existing point $v$ are consecutive, and thus can be added together
without locks.

\myparagraph{Optimization: Batch Size Truncation}
While allowing each point to see an index that is
roughly half the size it sees in the sequential setting,
prefix-doubling may still lose significant information in the
last few rounds when the batches are very large. To avoid this, we
upper bound the batch size by $\batchupperbound$, which we empirically
set to $0.02n$. This relaxation does not affect parallelism or
scalability in practice; for large datasets, 2\% of the input is more
than enough to utilize all threads on modern multi-core computers.
With this optimization, our prefix-doubling index achieved similar
quality as the sequential version: \pardiskann{} with $R=64,L=128$ on
a benchmark dataset \BIGANN-1M differs within 1\% of the QPS from the sequentially-built index, at the
same level of recall.


\subsection{{Clustering-Based Algorithms}}
\label{sec:general:tree}
Another approach for building an ANN graph is to use \emp{clustering
trees}.
At a high-level, the algorithm splits the input into two pieces,
and keeps recursively splitting until the number of points drops below
a given threshold, reaching a {\em leaf cluster}.
The structure of splitting points form a tree-like structure, called
a {\em cluster tree}.
The splitting step usually involves randomization, e.g., we can
generate a random hyperplane and split points based on which side of
the plane they fall.
Within each of the leaf clusters, a local ANN graph with stronger
conditions (e.g., connecting each point with some exact nearest
neighbors) is built.

Using different random seeds to generate different cluster trees, we
can generate multiple (overlapping) local ANN graphs.
The overall algorithm will obtain an ANN graph as the union of all
local ANN graphs, and obtains the final ANN graph by performing some
postprocessing.
These algorithms differ in the methodology in generating the
clustering tree, building the local ANN graphs, and/or postprocessing.
Among the algorithms in this paper, \hcnng{} and \pynn{} use the
clustering trees.

\myparagraph{{Challenges for Clustering-Based Algorithms}}
There are several challenges to efficiently construct ANN graphs in
parallel using this approach.
Firstly, some existing systems achieve parallelism simply by
parallelizing the construction of the $T$ trees (each tree is
constructed sequentially).
Since empirically the best value of $T$ is tens of trees (e.g., about
30 for \hcnng{})~\cite{munoz2019hcnng}, the algorithm naturally cannot
scale to more than $T$ threads in the tree construction step, which is
also the main reason that the original \hcnng{} implementation in
\cref{fig:buildparallelism} does not improve beyond 30 threads.
Secondly, existing parallel implementations also take per-point locks when
merging the edges from all the local ANN graphs, which causes
contention and non-determinism if pruning is used.
Lastly, some subroutines, such as the local ANN graph construction,
can generate costly (in terms of time or space) local structures,
which can become a performance bottleneck when the data size or the
number of threads is large.

Next, we present our general ideas to achieve better parallelism for
clustering trees.  In \cref{sec:hcnng,sec:pynn}, we further
discuss our new ideas to address the scalability issue in \hcnng{} and
\pynn{}.


\myparagraph{Parallelizing Clustering-Based Algorithms}
To parallelize the clustering-based algorithms, we apply two general
ideas.  First, we parallelize the construction of \emph{each
clustering tree}.
We then use parallel divide-and-conquer to always deal with both
branches in parallel, and use a parallel partitioning
primitive~\cite{blelloch2020toolkit,JaJa92} to assign points to
different branches in parallel.
This approach offers abundant parallelism across \emph{all leaves},
instead of just over the trees.
Although this is a natural idea, exposing more parallelism causes some
challenges, e.g., for \hcnng{}, more threads running in parallel causes
some space issues which we explain more in \cref{sec:hcnng}.

The second general technique is to avoid per-point lock when combining
edges in all local ANN graphs.  Instead of adding all edges
concurrently, our idea is to collect all edges in an array and run a
\emp{semisort} on it (see \cref{sec:prelims}), such that the edges
incident the same point are consecutive. The graph can be built
accordingly.

\begin{algorithm2e}[t]
	\caption{batchBuild($\mP, s, R, L$).\protect}
	\label{algo: batchBuild}\small
	\SetKwBlock{ParDo}{do in parallel}{end}
	\SetKwFor{ParFor}{parallel for}{do}{end}
	\SetAlgoLined
    \DontPrintSemicolon
    \SetKwProg{myfunc}{Function}{}{}
	\KwIn{Point set $\mP$, starting point $s$, beam width $L$, degree bound $R$.}
	\KwOut{An ANN graph consisting of all points in $\mP$.}
	$\mathtext{start} \gets 1$ \\
    \While(\tcp*[f]{Prefix-doubling}){$\mathtext{start}\leq|\mP|$}{
      $\mathtext{end}\gets \min(\mathtext{start}\times 2, \mathtext{start} + \batchupperbound, |\mP|)$\tcp*[f]{$\batchupperbound$: batch size upper bound}\\
      BatchInsert$(\mP[\mathtext{start}..\mathtext{end}])$\\
      $\mathtext{start} \gets \mathtext{end}+1$
    }
    \codeskip
    \myfunc(\tcp*[f]{Insert a batch $\mP'$ to the current index}){\upshape BatchInsert$(\mP')$}{
    	\ParFor{$p\in \mP'$\label{line:build:for1:start}}{
			$\mV, \mathcal{K} \gets \text{greedySearch}(p, s, L,1)$ \;
			$\Nout(p) \gets \text{prune}(p,\mV,R)$ \;\label{line:build:for1:end}
		}
		$\mathcal{B} \; \la \; \bigcup_{p\in \mP'} \Nout(p)$ \tcp*[f]{All (existing) affected points}\\
		\ParFor{$b \in \mathcal{B}$\label{line:build:for2:start}}{
            \tcp{$\mathcal{N}$: all points in $\mP'$ that added $b$ as their neighbors}
			$\mathcal{N} \gets \{p \; | \; p\in \mP' \; \land \; b \in \Nout(p)\}$ \;
			$\Nout(b) \gets \Nout(b) \cup \mathcal{N}$ \;
			\lIf{$|\Nout(b)| > R$}{
				$\Nout(b) \gets \text{prune}(b,\Nout(b),R)$
			}
			\label{line:build:for2:end}
		}
    }
\end{algorithm2e}

\section{\sysname{} Algorithms}\label{sec:algos}

In this section, we further describe
four graph-based ANNS algorithms that benefit from our
techniques proposed in \cref{sec:algsurveyed}.
In addition to the general techniques,
we also employ specific optimizations for each individual algorithm
to improve their scalability, which will be introduced below.
%

\subsection{DiskANN}
\label{sec:diskann}
\diskann{}~\cite{subramanya2019diskann} is a system consisting of an
incremental in-memory ANNS graph algorithm as well as a system for
storing the graph on an SSD. We focus on only the incremental ANNS
graph algorithm as our work focus on the in-memory ANNS system.
The in-memory \diskann{} algorithm is almost completely described by
\cref{algo: insert}, with the exception of the pruning step.
In the paper on the navigating spreading-out graph
(NSG)~\cite{fu2019nsg}, Fu et al.~proposed a pruning method on the
visited list $\mV$: roughly, they repeatedly select the point $p^*$
closest to $p$ in $\mV$, then filter out points $p'$ that are
\textit{($\alpha$ times) closer to $p^*$ than to $p$} (i.e.,
remove all $p'$ s.t. $\alpha \dist{p^{*}}{p'} \leq \dist{p}{p'}$).
%
This can be thought of as streamlining navigation by pruning out long edges of triangles.
%
%
As this technique is general, we also apply the $\alpha$ parameter to
other algorithms in this paper to reduce their degrees
(and thus make the ANN graph sparser) when possible, in order to make a more fair comparison.


To adapt \diskann{} for machines to be scalable to hundreds of cores
in the in-memory setting, we used the prefix-doubling approach as
described in the previous section.

\subsection{HNSW}\label{sec:hnsw}
The hierarchical navigable small world (\hnsw{})
algorithm~\cite{malkov2020hnsw} is an incremental algorithm that
constructs a hierarchical structure (intuitively the structure is
similar to a skip list);
each layer of the hierarchy is a navigable
small world (NSW) graph~\cite{ponomarenko2011approximate}.
In a NSW graph, nodes tend to be connected to their near neighbors,
while ensuring that the overall graph is {\em navigable}, i.e., a
search can reach any node in a small number of hops.
%

\hnsw{} builds multiple layers of NSW graphs so that the lower layers
are supersets of the upper layers.
The number of vertices in each layer increases geometrically from top to
bottom, and the bottom layer contains all the input points
(conceptually this is similar to a skip list).
Insertion in an NSW graph is also similar to ~\cref{algo: insert}. The
\texttt{prune} scheme in \hnsw{} is similar to \diskann{} in that it
prunes out long edges of triangles,
but also includes some additional heuristics.

For search, \hnsw{} traverses through the layers one at a time. It
starts at the top layer, looks for the $1$-nearest neighbor $p$ of
the query point using \cref{algo: greedySearch} with beam size 1, and
shifts down to the next layer at $p$ to repeat the procedure until
reaches the last layer.
Then, taking the current result as the entry point, it runs
\cref{algo: greedySearch} to obtain the $k$-nearest neighbors at the
bottom layer.

In our implementation (\parhnsw{}), we utilize parallel
prefix-doubling.  To adapt prefix-doubling to the the multi-level
hierarchical structure, we simply use batch insertion for each layer.
We also carefully remove locks in all internal data structures in \hnsw{}.

\subsection{HCNNG}\label{sec:hcnng}
The hierarchical clustering-based nearest neighbor graph
(\hcnng{})~\cite{munoz2019hcnng} uses the clustering-based approach.
The clustering works by randomly selecting two points $p_1$ and $p_2$,
and partitioning the input by deciding whether a point is closer to
$p_1$ or $p_2$.
Leaf clusters are obtained when the number of points is below a given
threshold.
Within a leaf, the local ANN graph is a degree-bounded minimum
spanning tree (MST), i.e., an MST where each point has degree at most
$K$.
Pruning is then applied to remove redundant edges.

\myparagraph{Reducing Work and Space using Edge-Restricted MSTs}
We parallelized \hcnng{} without locks by constructing the clustering
trees and merging edges in parallel as mentioned in
\cref{sec:general:tree}.
However, extra challenges emerge when a large number of threads can run
in parallel.
In particular, the MST is of the {\em complete graph} containing all
pairwise distances of points in a leaf.
When hundreds of threads perform this process on different leaves in
parallel, the temporary memory usage can be very high.
In our experience, storing all pairwise edges exceeds the L3 cache on
our machines, and severely limited speedup.  To remedy this, instead
of building the MST over all potential edges, we build an
\textit{edge-restricted MST}: instead of generating all pairwise
edges, the MST is based on a graph where each point is connected with
its $l$-nearest neighbors for some small $l$ (we use 10).
This optimization significantly saved space and in turn improved
parallelism with no drop in QPS for a given recall.
Our \parhcnng{} is up to 12$\times$ faster than the original \hcnng{}
implementation (see \cref{fig:buildparallelism}), and achieves good
self-relative speedup.

\subsection{PyNNDescent}\label{sec:pynn}
The \pynn{}~\cite{mcinnes2020pynndescent} algorithm uses a
combination of a clustering-based approach to find an initial set of
out edges along with iterative refinement to improve the set.
The clustering initially used to construct the graph is based on
choosing random hyperplanes.
The local ANN graphs connects each point to the exact $K$ nearest
neighbors within each leaf.
In addition to the clustering-based approach, \pynn{} also includes
a special postprocessing called \emph{nearest neighbor descent}, which runs
in an iterative way.
Each round begins by undirecting the graph, i.e., adding the opposite
edge of each directed edge.  Then, each point $p$ computes its two-hop
neighborhood $\mathcal{Q}$ and retains the $K$ closest candidates
among the points $q \in \mathcal{Q}$.
The algorithm terminates once only a small fraction of edges change on
each round (i.e., converges).
We then use a pruning algorithm to prune out the long edges of all
triangles.

\myparagraph{Optimizing Parallelism and Random Edge Sampling}
We had to significantly modify the \pynn{} algorithm to scale to
large datasets, and indeed as shown in \cref{sec: results},
despite our optimization efforts we were not able to scale \pynn{} to
datasets with billions of points.
However, our techniques still make it achieve reasonable QPS and recall on
inputs with $\sim$100 million points.

The fundamental challenge is that calculating the neighbors of
neighbors of a vertex requires work (and space) proportional to the
square of the degree.  We used two ideas to address this challenge.
First, note that undirecting the graph edges can significantly
increase the degree of a vertex.
Thus, in edge undirecting, we limit each vertex's degree to be
at most $2000$ by randomly sampling edges, which makes the quadratic work more manageable.
Also, we compute sets of two-hop neighborhoods in batches
rather than all at once (i.e., we limit parallelism to limit the
amount of intermediate memory used).
With these optimizations, we were able to make our implementation, \parpynn{},
scale to 100M points, but the amount of temporary memory required to
store two-hop graph made it infeasible to
scale 
to a billion points.

\subsection{Search and Layout Optimizations}
\label{sec:query-optimization}

In our experiments we use the {\em same} beam search algorithm across
all of our implementations of \pardiskann{}, \parhcnng{} and
\parpynn{} since they all generate a graph in the same format.
The only difference is in how we select a start vertex.
Our search algorithm for \parhnsw{} is also very similar, but slightly
different since it needs to move between levels of the hierarchy.
We have made a handful of modest optimizations to the search for all
algorithms over the generic form given in \cref{algo: greedySearch},
which we describe here.

Firstly we use an optimized approximate hash table with one-sided
errors to quickly identify whether a point is in the visited
set $\mV$.
Each point is inserted to the hash table by finding a random position.
%
When two vertices map to the same position, only one will be stored,
and the second will be revisited if encountered.
The table size is set as the square of the beam size, which is
large enough that revisiting is rare but is small enough to fit the
table in the first-level cache.
This is especially useful for improving the performance of the
original \hnsw{}, where a per-point flag array is used to check
membership in $\mV$, and in general improved the performance for all
our algorithms by 28.6\%--44.5\%.

We also avoid levels of indirection in the graph layout.
In particular the edge-list for each vertex is kept at a fixed length
so we can calculate its offset from the vertex id.
We also use an $(1 + \epsilon)$ pruning during the search as suggested by
Iwasaki and Miyazaki~\cite{iwasaki2018optimization}.
In particular we only search vertices which have a distance to the
search point that are within a factor of $(1 + \epsilon)$ of the
current $k$-th nearest neighbor.
The $\epsilon$ is tuned based on the desired accuracy, but is never
greater than .25.
When sweeping the query parameters to obtain different points on the
QPS/recall tradeoff curve, we therefore sweep two parameters: the beam
size and $\epsilon$.


\section{Experimental Evaluations}\label{sec: results}
In this section, we evaluate \sysname{} and present interesting
findings from experiments at the end.
We implement \sysname{} using C++ using
ParlayLib~\cite{blelloch2020toolkit} to support fork-join parallelism.
We also use some standard building blocks (e.g., sorting, semisorting,
partition) from ParlayLib.

\subsection{Experimental Setup}\label{sec: experimental}

\myparagraph{{Datasets}} We utilize three billion-size datasets for the
majority of our experiments; we accessed these datasets through the
BigANN Benchmarks competition framework, and some of these datasets
were released for the competition~\cite{simhadri2021results}. The
widely used \emph{\BIGANN{} dataset}\footnote{Note that throughout the
paper we use BigANN to refer to the benchmarking framework, and
\BIGANN{} to refer to the dataset.} consists of SIFT image similarity
descriptors applied to images~\cite{simhadri2021results,
jegou2011searching, douze2011product}. It is encoded as
128-dimensional vectors using 1 byte per vector entry.
The \emph{Microsoft SPACEV dataset (\SPACEV)} encodes web documents
and web queries sourced from Bing using the Microsoft SpaceV Superior
Model. The goal is to match web queries with appropriate web
documents; the dataset consists of 1 byte signed integers in 100
dimensions~\cite{spacev2021spacev}. The \emph{Text2Image dataset
(\TTI)}, released by Yandex Research, consists of a set of images
embedded using the SeResNext-101 model, and a set of textual queries
embedded using a DSSM model. Its vectors are represented using 4 byte
floats in 200 dimensions~\cite{baranchuk2021benchmarks}.
%

\myparagraph{{Machines}} For most experiments, we used an
AWS c6i-series virtual machine with two 3rd Generation Intel® Xeon® Gold Processors with 128 vCPUs available to the user, and 1 TB main memory.
%

For the billion-scale results on \TTI{}, we used an AWS x2idn-series virtual machine with two 3rd Generation Intel® Xeon® Platinum Processors with 128 vCPUs available to the user, and 2 TB main memory.

For Figure~\ref{fig:buildparallelism} we used an AWS c7i-series virtual machine with one 4th Generation Intel® Xeon® Gold Processor with 96 vCPUs available to the user, and 192 GiB main memory.

\myparagraph{Measurement}
We report build times and QPS using all threads
unless stated otherwise; throughout the experiments, we use QPS as
opposed to latency, since QPS is more relevant to large multicore
machines, and algorithms are typically always within an acceptable
latency range.
As discussed in \cref{sec: intro}, ANNS algorithms are primarily
evaluated based on the {\em recall-QPS} curve, i.e., a curve where
the $y$-axis is the QPS and the $x$-axis is the recall. To obtain
points on this tradeoff curve, we perform a parameter sweep. Typically
this is done by building a single (fixed) index, and then adjusting
the parameters for a search, e.g., the beam-width, and $\epsilon$
value.

\myparagraph{Baseline Algorithms} We compare all our implementations
with the original implementations of
\diskann{}~\cite{subramanya2019diskann},
\hnsw{}~\cite{malkov2020hnsw}, \hcnng{}~\cite{munoz2019hcnng}, and
\pynn{}~\cite{mcinnes2020pynndescent}, on the 1M-scale \BIGANN{}
dataset to demonstrate the improvement in scalability and parallelism
over the existing implementations.
The baseline implementations are carefully chosen from the BigANN
benchmark to select the most competitive existing algorithms.
The original \hnsw{} implementation is safe for concurrent operations
due to using locks, but does not exploit parallelism by default.
%
We added a batch-parallel interface to the original \hnsw{} using ParlayLib.
For larger scale experiments, we compare \sysname{} to two non-graph
algorithms based on inverted indexing (IVF): \faiss{} and
\falconn{}.
For completeness, we describe these two algorithms in the
supplemental material, and provide a more complete list of algorithms
that we did not include in the study, along with the reasons for their
exclusion.

\myparagraph{{Algorithm Parameters}}
Our interest is in optimizing for the high recall regime (from .9 to
.999) at the highest QPS possible.
For reproducibility, we provide our choices of parameters in the
supplemental material, which are chosen to give the best performance
based on both our own experiments and the literature.
\hide{
For \diskann{}, \hnsw{}, \hcnng{}, and \pynn{}, the parameters shown
are applied to all dataset sizes.
For \faiss{} and \falconn{}, the parameters are specific to the size
of the dataset.
\yihan{I hide this because I didn't see specific parameters about faiss and falconn given based on the size of the dataset in the appendix.}
}

\myparagraph{Code Availability} Our source code is available at \url{https://github.com/cmuparlay/ParlayANN}.

\hide{
\myparagraph{{Search Parameters}} In all cases, we test a wide variety
of search parameters and report those that maximize QPS for a given
recall.
For our graph-based algorithms, which all use beam search, we vary $L$
(beam width),
and $\epsilon$ (the discard
parameter) and report the best results for a variety of recall
thresholds.
For \faiss{}, we vary the number of cells probed as well as the
\hnsw{} \emph{efConstruction} parameter (for \faiss{} settings which use
\hnsw{} as a sub-quantizer) and report the best results.
For \falconn{}, we vary both the number of probes and the maximum
number of candidates in a wide range.
\yihan{the parameters look different from what was mentioned in section 3.}
\laxman{I think we can potentially drop this entire paragraph or push
it to the supplementary materials. I think there was some
inconsistency in terminology which I hope I fixed (i.e., using Q and c
here, but L and epsillon in Section 3.}
}


\begin{figure*}[t]
	\begin{subfigure}[h]{\textwidth}
		\includegraphics[width=.56\textwidth]{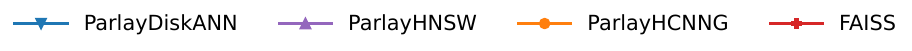}
		\centering
		\vspace{-.5em}
	\end{subfigure}
	\begin{subfigure}[b]{.31\textwidth}
		\centering
		\includegraphics[width=\textwidth]{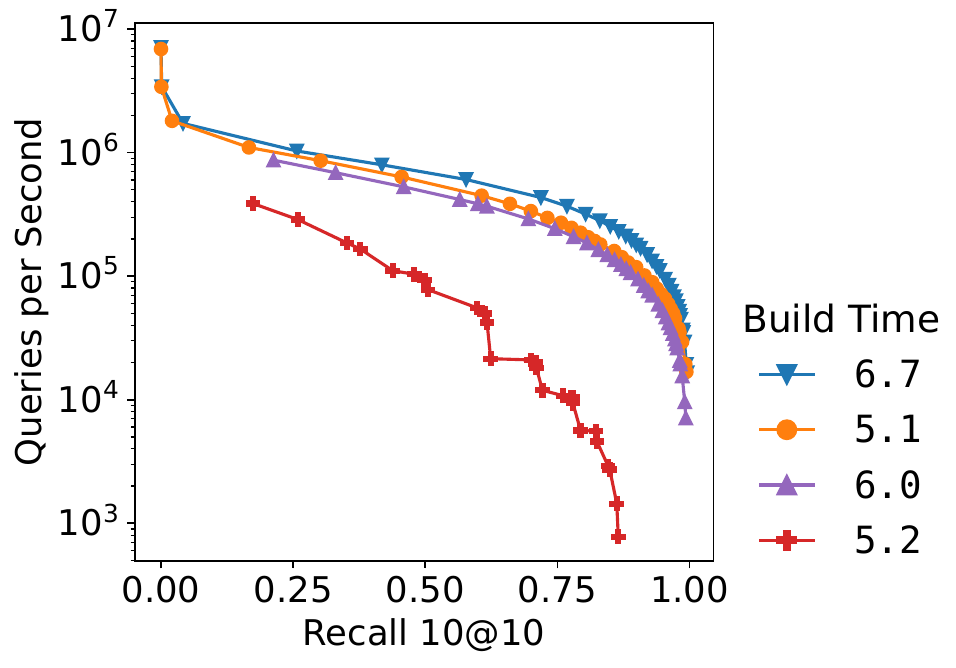}
        \caption{\BIGANN-1B QPS}\label{fig:bigannqps}
	\end{subfigure}
	\hfil
	\begin{subfigure}[b]{.31\textwidth}
		\centering
		\includegraphics[width=\textwidth]{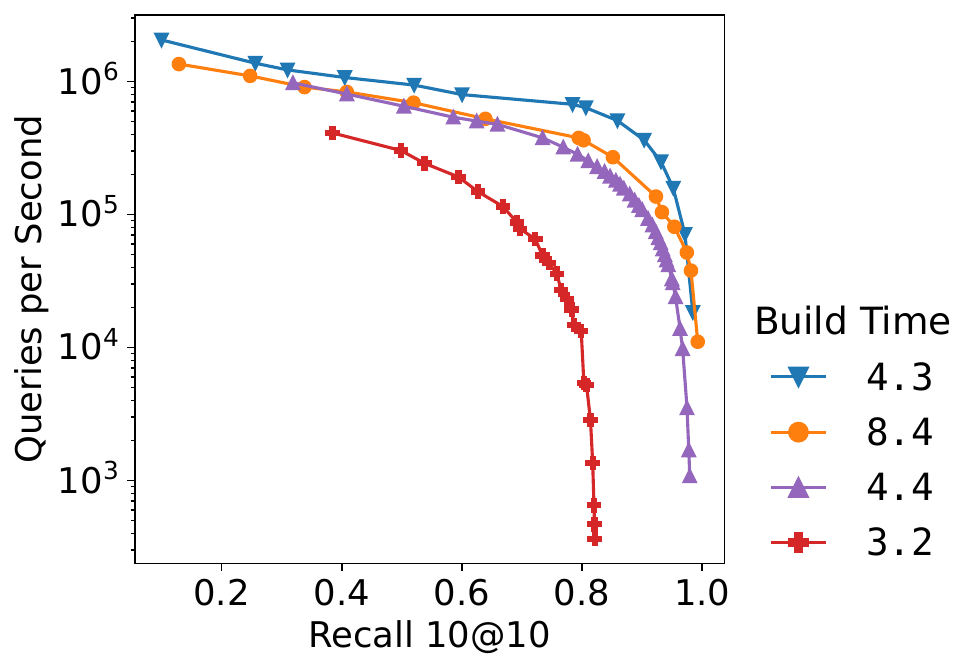}
		\caption{ \SPACEV-1B QPS}\label{fig:spacevqps}
	\end{subfigure}
	\hfil
	\begin{subfigure}[b]{.31\textwidth}
		\centering
		\includegraphics[width=\textwidth]{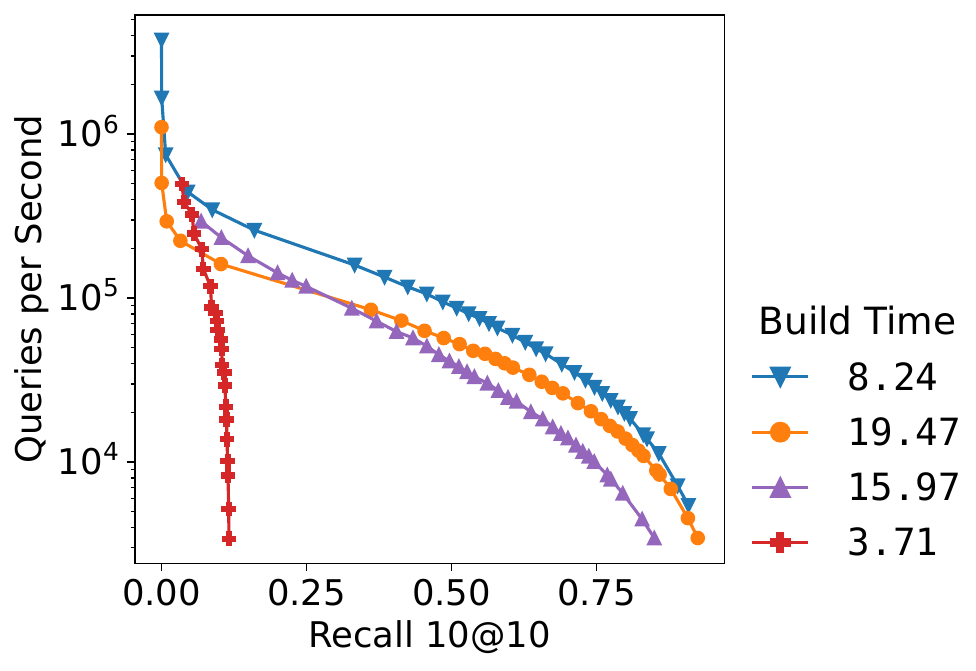}
		\caption{ \TTI-1B QPS}\label{fig:oodqps}
	
	\end{subfigure}
	\vspace{1em}

	\begin{subfigure}[b]{.31\textwidth}
		\centering
		\includegraphics[width=\textwidth]{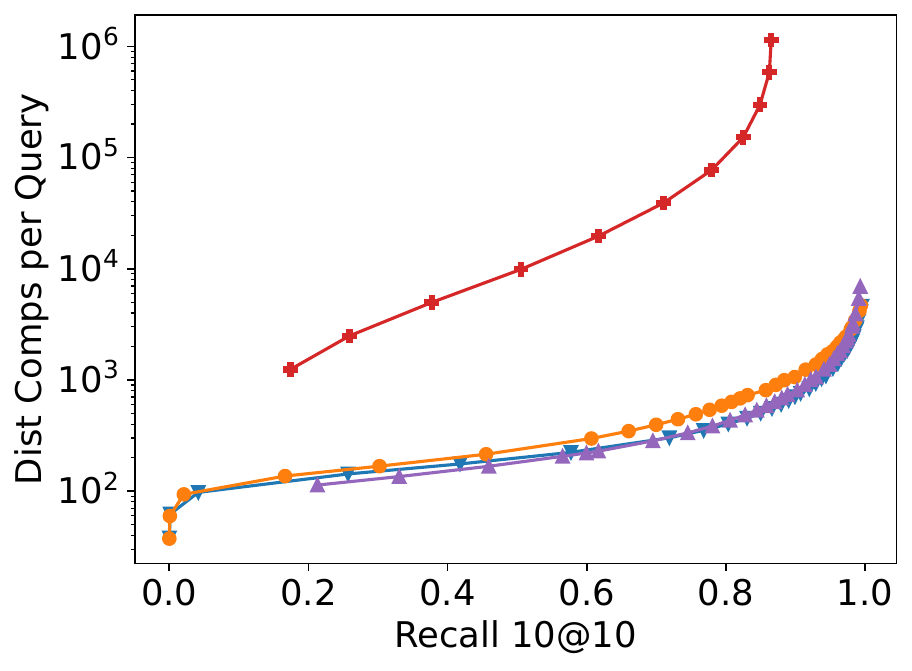}
		\caption{ \BIGANN-1B Dist Comps}\label{fig:biganndistcmps}
	\end{subfigure}
	\hfill
	\begin{subfigure}[b]{.31\textwidth}
		\centering
		\includegraphics[width=\textwidth]{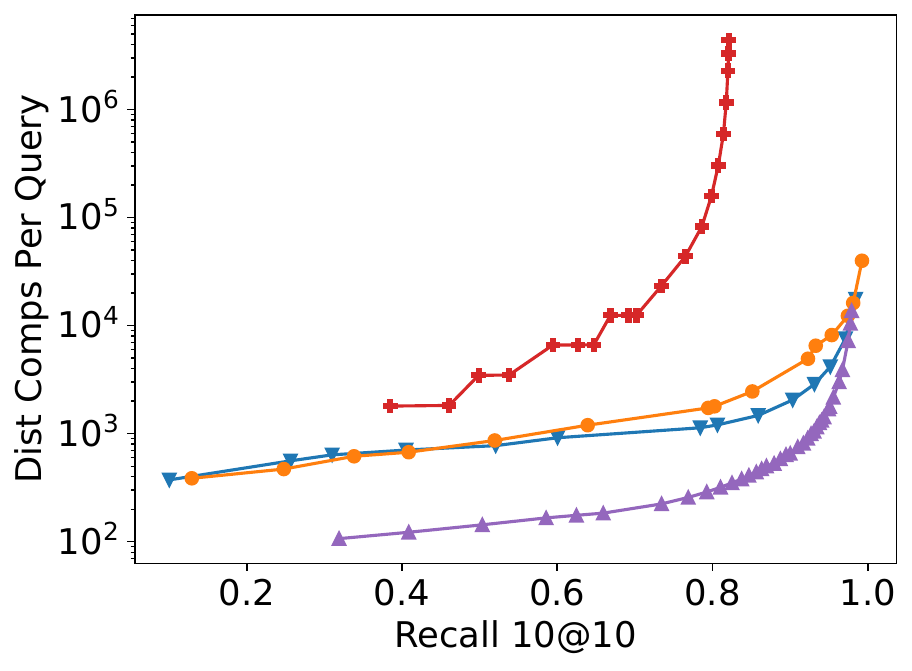}
		\caption{ \SPACEV-1B Dist Comps}\label{fig:spacevdistcmps}
	\end{subfigure}
	\hfill
	\begin{subfigure}[b]{.31\textwidth}
		\centering
		\includegraphics[width=\textwidth]{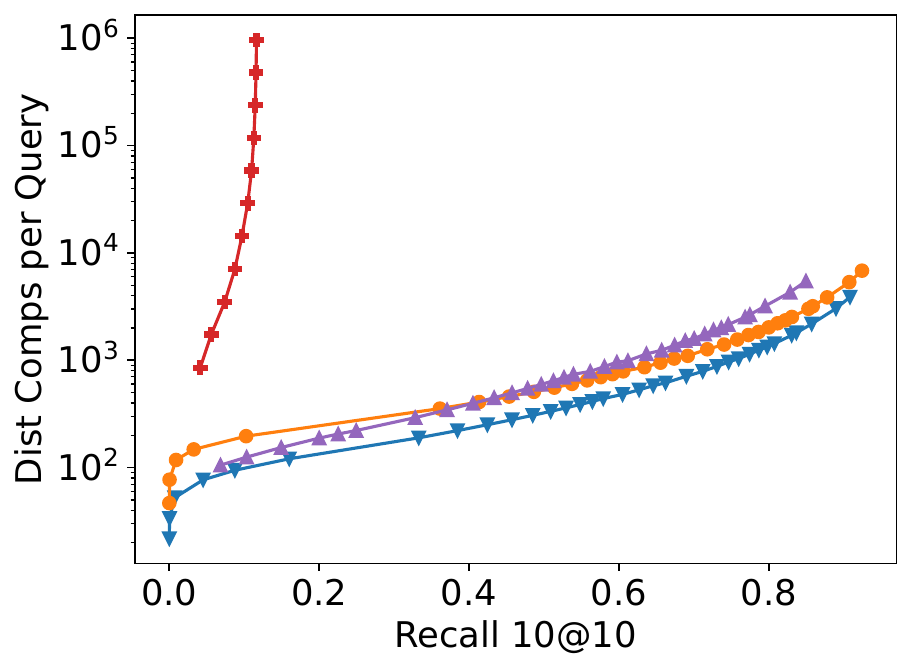}
		\caption{ \TTI-1B Dist Comps}\label{fig:ooddistcmps}
	\end{subfigure}
	\vspace{1em}
	\caption{Build time (hours), QPS, recall, and distance comparisons for all algorithms on billion-size datasets. 
}
\label{fig:billionsize}

\end{figure*}

\begin{figure*}[t]
	\hfill	
	\begin{subfigure}[b]{.3\textwidth}
		\centering
		\includegraphics[width=\textwidth]{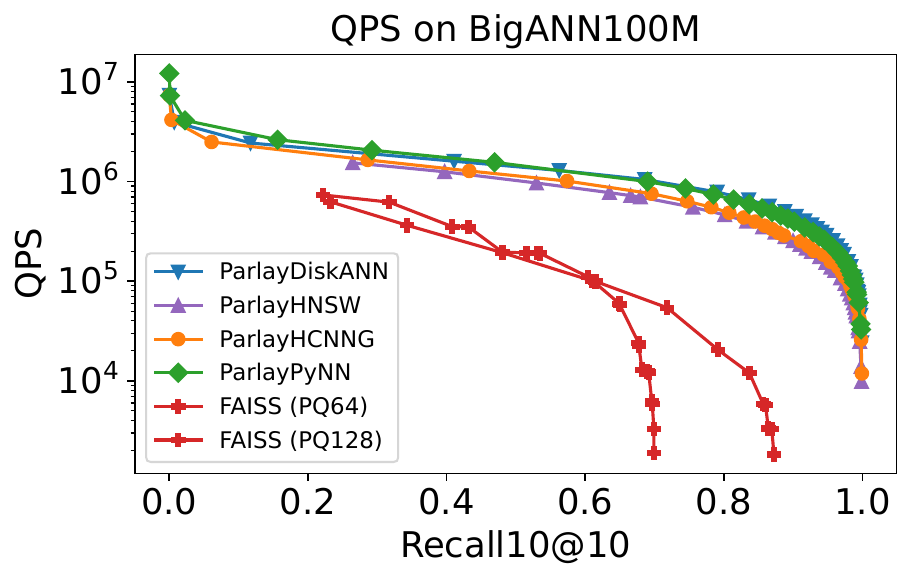}
		\caption{\BIGANN-100M}\label{fig:bigannQPSEv5}
	\end{subfigure}
	\hfill
	\begin{subfigure}[b]{.3\textwidth}
		\centering
		\includegraphics[width=\textwidth]{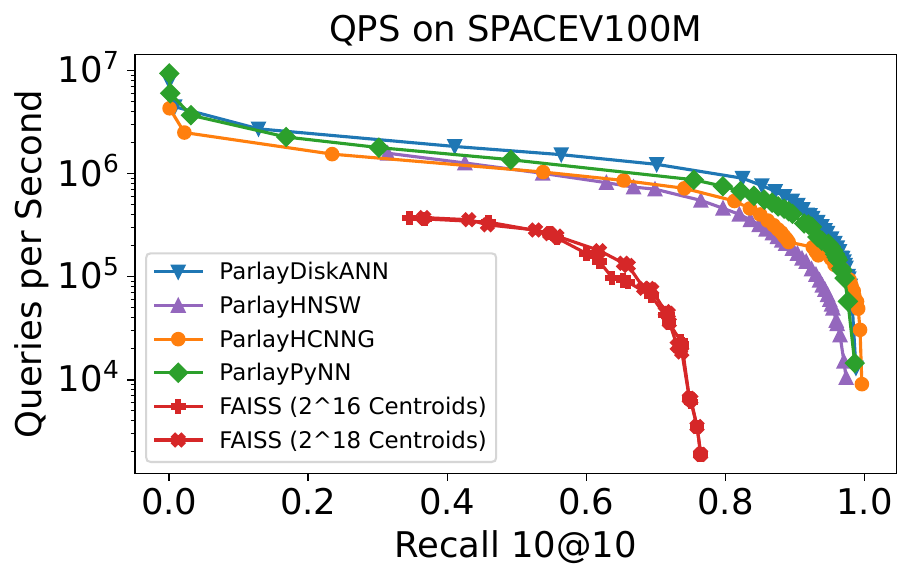}
		\caption{\SPACEV-100M}\label{fig:spacevQPSEv5}
	\end{subfigure}
	\hfill
	\begin{subfigure}[b]{.3\textwidth}
		\centering
		\includegraphics[width=\textwidth]{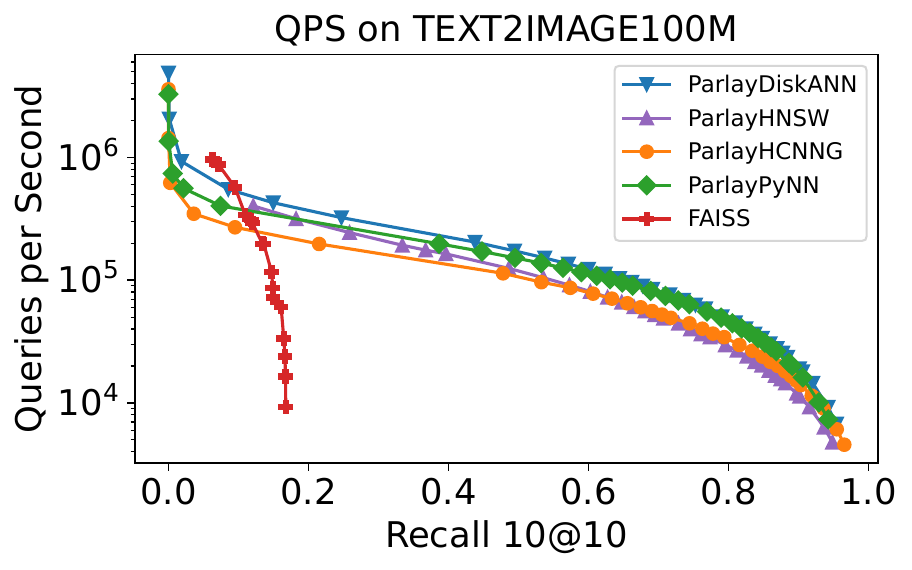}
		\caption{\TTI-100M}\label{fig:text2imageQPSEv5}
	\end{subfigure}
	\hfill
	\\
	\hfill
	\begin{subfigure}[b]{.3\textwidth}
		\centering
		\includegraphics[width=\textwidth]{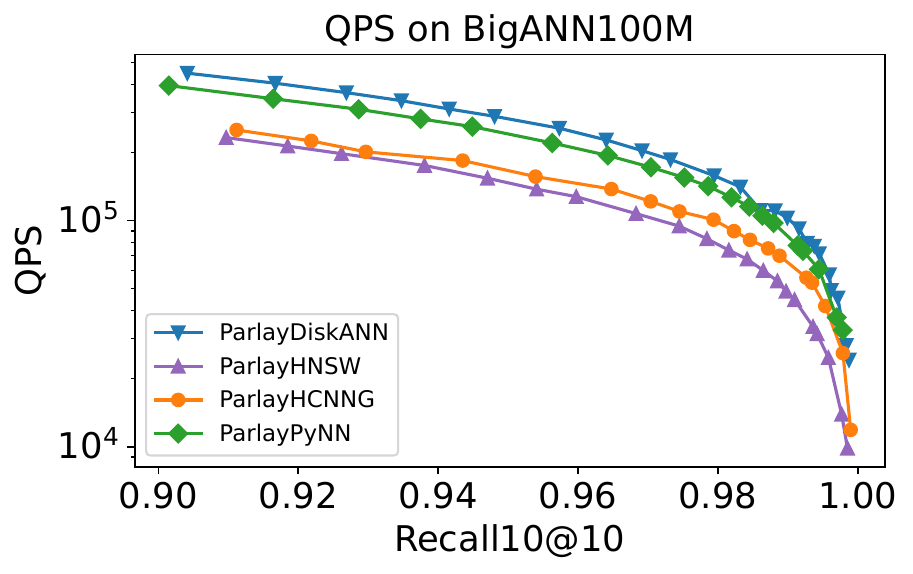}
		\caption{\BIGANN-100M }\label{fig:bigannQPSEv5High}
	\end{subfigure}
	\hfill
	\begin{subfigure}[b]{.3\textwidth}
		\centering
		\includegraphics[width=\textwidth]{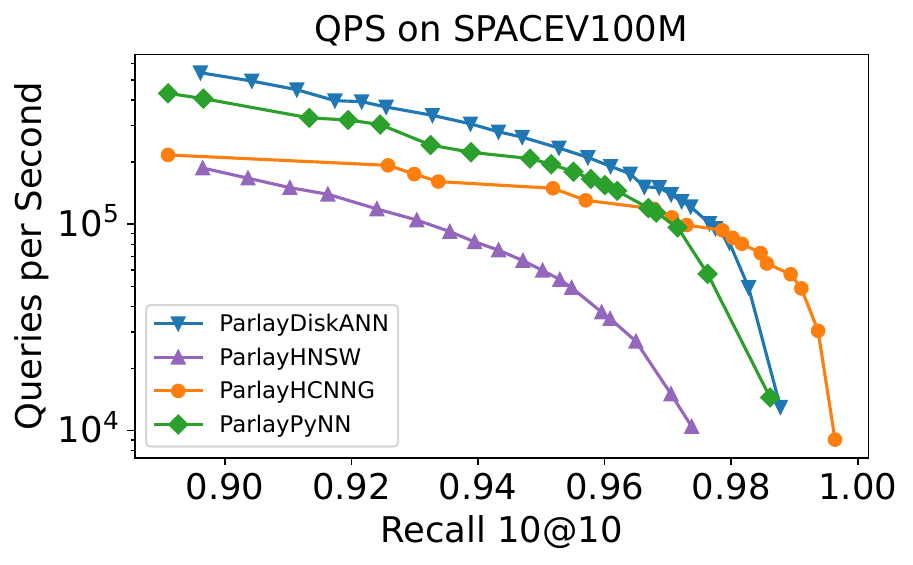}
		\caption{\SPACEV-100M }\label{fig:spacevQPSEv5High}
	\end{subfigure}
	\hfill
	\begin{subfigure}[b]{.3\textwidth}
		\centering
		\includegraphics[width=\textwidth]{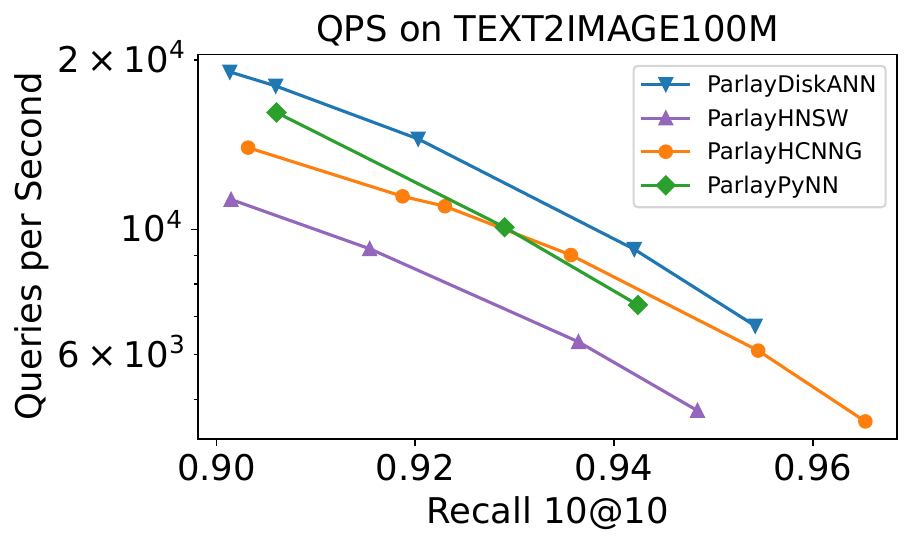}
		\caption{\TTI-100M}\label{fig:qpst2ihigh}
	\end{subfigure}
	\hfill

	\caption{QPS-recall curves on all 100-million size datasets. The first row shows the overall QPS/recall curve, while the second row zooms into a higher-recall regime. The build times are given in \cref{fig:buildtimeshundredmil}
	}\label{fig:100Mresults}
\end{figure*}

\begin{table}[t]
	\begin{center}
		\begin{tabular}{c@{  }@{  }c@{  }@{  }c@{  }@{  }c@{  }@{  }c}
			& \BIGANN & \SPACEV & \TTI  \\
			\hline
			DiskANN & .42 & .35 & .70  \\
			HNSW & .35 & .37 & .94  \\
			HCNNG & .45  & .77  & 1.75   \\
			pyNNDescent & .42  & .73  & 1.23  \\
			FAISS & .19 & .13 & .22  \\
			\hline
		\end{tabular}
	\end{center}
	\caption{ Build times (hours) on hundred million scale datasets.}\label{fig:buildtimeshundredmil}
\end{table}



\subsection{Comparison with ANN Benchmarks}

\begin{figure}[t]
	\centering
	\includegraphics[width=.95\columnwidth]{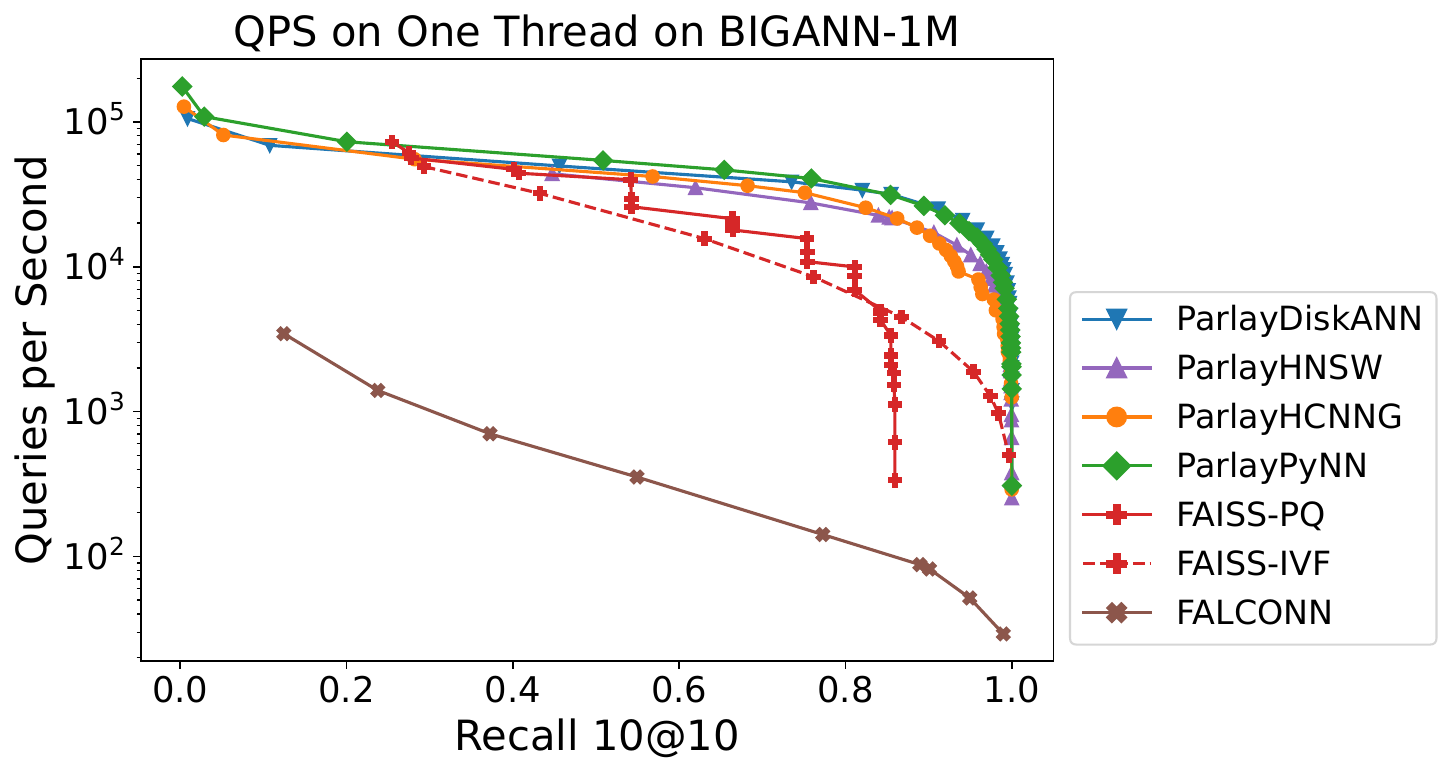}
	\caption{ QPS on a single thread on \BIGANN-1M. Shown to compare with ANN-benchmarks.\label{fig:baseline}}
\end{figure}

First of all, we demonstrate the single thread performance of \sysname{} on \BIGANN-1M in \cref{fig:baseline}. We refer to the parameter settings in the ANN Benchmarks framework~\cite{amuller2020ann}, and compare to the publicly-available numbers on the website.
The single-thread performance of \sysname{} roughly match the results on ANN Benchmarks website~\cite{annbench}. Due to \falconn's poor performance \BIGANN-1M, and its correspondingly low performance on the hundred million and billion size datasets, we do not include \falconn{} in further figures. 

\subsection{Parallelism and Scalability}
To substantiate our claims of improving the parallelism of each
graph-based algorithm as well as illustrate issues with the
parallelism of the original implementations, we compare \sysname{} with
the original implementations of each algorithm.
We present the performance of building the index (graph) as the number
of threads increases in \cref{fig:buildparallelism}.
\hide{ (we measure this
in terms of {\em work} or the running time multiplied by the number of
threads; an algorithm with perfect self-speedup will have a flat
curve).}
For the same algorithm, all numbers (both original and ours) presented
are \emp{running time speedup relative to the original implementation on
one core}.
Therefore, the curve provides a direct running time comparison between the original
implementation and our implementation (higher is better).
For each algorithm, the two implementations always use the same
parameters, and achieve similar query quality (except for some where
\sysname{} also improved queries and achieved \emph{better}
query quality).

For \diskann{}, we find about 1.2$\times$ improvement in performance
by \sysname{}.
The original \diskann{} scales well to 30 to 60 threads but
eventually the use of locks leads to performance degradation on more
threads.
\hnsw{} suffers from similar locking-related issues, and \parhnsw{}
performs much better with more than 50 threads, and eventually
achieves 1.4$\times$ better performance.
As mentioned in \cref{sec:algsurveyed}, the original \hcnng{} only
exploits parallelism by building all clustering trees in parallel, and
fails to scale beyond $T$ threads as a result.
Our \parhcnng{} was both faster on a single thread and even better
when the number of threads increases, and eventually becomes 12$\times$ faster
than their implementation when using all threads.
PyNNDescent's original implementation used Numba~\cite{lam2015numba}
for parallelism and did not scale beyond 16 threads on our machine.
Our implementation eventually becomes 28$\times$ faster than their
parallel implementation.
%
\hide{All our implementations achieved speedup up to all 96 cores; three
of them further benefit from hyperthreading.
\laxman{I think we can drop this last sentence or add numbers to it.}
}

\begin{figure*}[h]
	\begin{subfigure}[h]{\textwidth}
		\includegraphics[width=0.68\textwidth]{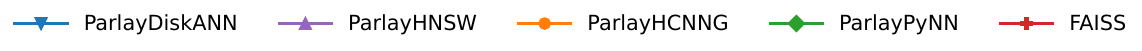}
		\centering
		\vspace{-.5em}
	\end{subfigure}
	\begin{subfigure}[b]{.31\textwidth}
		\centering
		\includegraphics[width=\textwidth]{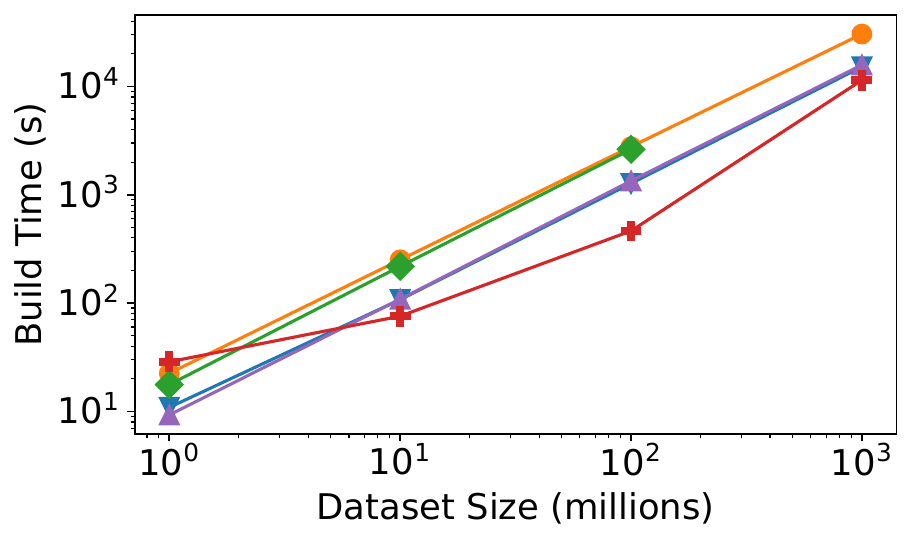}
		\vspace{-2em}
		\caption{ Build times shown on \SPACEV{} as dataset size increases.}
		\label{fig:buildsscale}
	\end{subfigure}
	\hfill
	\begin{subfigure}[b]{.31\textwidth}
		\centering
		\includegraphics[width=\textwidth]{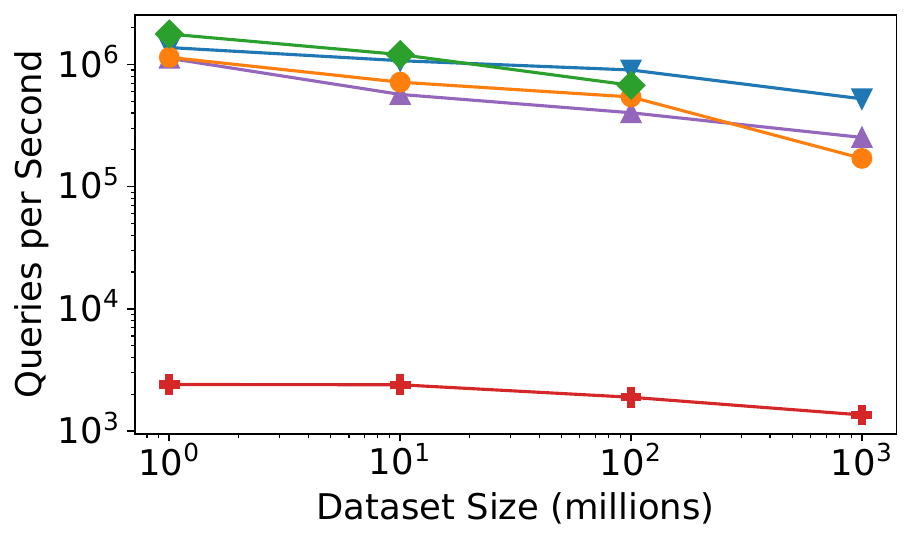}
		\vspace{-2em}
		\caption{ QPS for fixed recall (.8) on \SPACEV{} as dataset size increases.}\label{fig:qpsscale}
	\end{subfigure}
	\hfill
	\begin{subfigure}[b]{.31\textwidth}
		\centering
		\includegraphics[width=\textwidth]{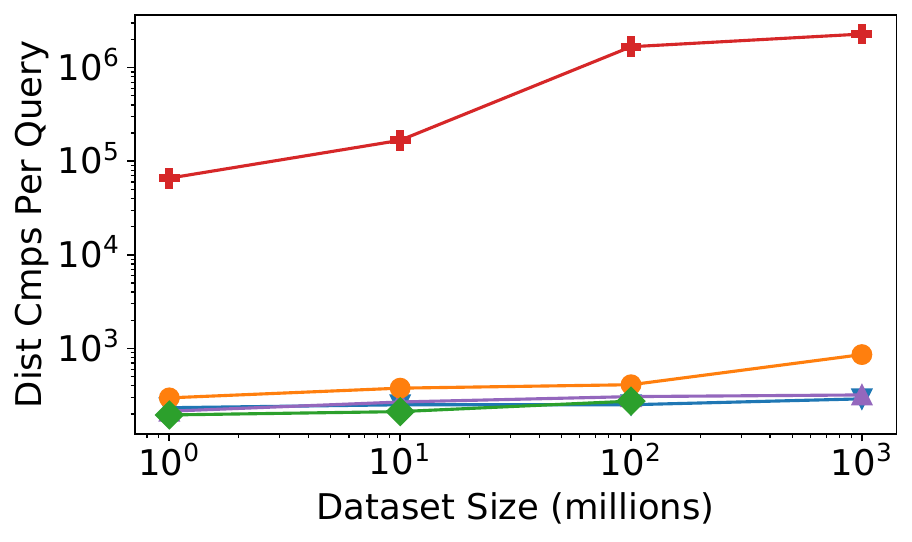}
		\vspace{-2em}
		\caption{ Dist computations per query for fixed recall (.8) on \SPACEV{} as dataset size increases.}\label{fig:distcmpsscale}
	\end{subfigure}
	\caption{ Figures showing the effect of dataset size on different metrics using the \SPACEV dataset.
}\label{fig:scaling}
\end{figure*}

\subsection{Full Billion-Scale and Hundred-Million Results}

In this section we present our results for all algorithms and
for three billion-scale datasets as well as their hundred-million scale versions.

\cref{fig:billionsize} shows the QPS-recall and distance-comparison-recall curves for all tested algorithms on the three billion-scale dataset,
along with the corresponding time to build their indexes presented on the side.
%
%
As mentioned in \cref{sec:algsurveyed}, \parpynn{} is not present in
the billion-scale figures since its memory requirements were infeasible for
billion-scale datasets; it can be found in the hundred million-scale
experiments. It is competitive with the other algorithms at the hundred-million scale.

In general, all our graph-based implementations achieve similar performance in
both build and query.  All of them can build the billion-scale indexes
in around 10 hours.
Among them, \parhnsw{} has slightly shorter build time (up to
2.3$\times$ faster than the other two), and \pardiskann{} is slightly
better in query (the recall-QPS curve is almost always at the top).

The non-graph algorithms we compared to achieved faster index
building time, where \faiss{} is usually 1.5--3$\times$ faster than the
graph-based algorithms.
However, both of them (especially \falconn) \emph{struggled to get
high recall on all datasets}\footnote{
We made many attempts to achieve the best query quality for \faiss{}
and \falconn{}, including increasing the building time and using
the suggested parameters from existing resources (e.g., FAISS
Wiki~\cite{faisswiki}).  The results we present are the best we
achieved after extensive experiments.}.

For \BIGANN{} and \SPACEV{}, \faiss{} did not achieve a recall higher than 0.8
even with very low QPS.
At 0.8 recall, \faiss{} has orders of magnitude lower QPS than
the graph-based algorithms (although at lower recall values, the gap between algorithms is significantly smaller).

\faiss{} achieves QPS close to (but still lower) the graph
algorithms at low recall values, but the QPS drops dramatically when a
recall higher than 0.6 is desired.

\faiss{} also performs especially poorly on the
out-of-distribution (OOD) dataset \TTI, where both of them only achieved
0.2 recall at most.

Ultimately, higher build times may be acceptable if the resulting
index can achieve high recall and QPS.
From this perspective, we find that the graph algorithms adapt better
to achieve high-recall and QPS on billion-scale datasets compared with
non-graph ones.
For \BIGANN{}, all of the three graph-based algorithms eventually can
achieve close to 100\% recall at about $10^4$ QPS.
For \SPACEV{}, \parhcnng{} achieves close to 100\% recall at $10^4$
QPS, while the other two can also achieve a recall above 0.9.
%


This advantage (high recall) of the graph-based algorithms is
especially true for queries that are out-of-distribution (OOD).
While the query quality of the non-graph algorithms seemed to be
severely affected by the OOD queries, all the three graph-based
algorithms were still capable of achieving a recall of 0.8 or more on
this challenging OOD dataset (\pardiskann{} can even achieve a recall
at 0.9).
At the same recall, the QPS of the graph-based algorithms is 12.2--19.6$\times$ slower compared to the other non-OOD datasets.

\subsection{Dataset Size Scaling}

How do ANNS algorithms scale as we increase the size of the dataset?
%
%
We start with the \SPACEV{} dataset as an example to explore this
question and present the result in \cref{fig:scaling} at a fixed
recall of 0.8.
In addition to build times and QPS, we also measure the average
distance computations per query for each algorithm.
We study this metric because for most ANNS algorithms on
high-dimensional points, the distance comparison are the most
expensive part.

For our graph-based algorithms, we found the build times incurred
slightly superlinear increases as the dataset size increased
(\cref{fig:buildsscale}); build times increased by a multiplicative
factor of 11--12$\times$ when the size of the dataset increased by 10$\times$.
For \parhnsw{} and \pardiskann{}, this superlinear increase can be
attributed to the mechanics of the beam search: on a larger graph,
beam search takes longer to terminate as there are more suitable
candidates in its frontier.
For \parpynn{}, we found that the nearest neighbor descent process
consistently took more rounds to terminate for larger dataset sizes.
Since the nearest neighbor graph for a larger dataset will likely have
a larger diameter, two-hop exploration takes longer to ``propagate''
through the entire graph.
For \faiss{}, we found an unusually small increase in build time
between the 10M and 100M datasets.
We attribute this to issues with parallelism that become less of a
bottleneck at higher numbers of data points.
\hide{(in the supplemental
material we show that FAISS does not achieve parallel speedup past 20
threads on \BIGANN-1M).
\yihan{did we do this? Which figure is this referring to?}
\laxman{let's check; if not I think we can drop this sentence without
hurting the story---it doesn't matter if FAISS scales well on large
datasets since the quality is not good for anything other than range
search}
}

For QPS (see \cref{fig:qpsscale}), \pardiskann{} and \parhnsw{} show a
steady decrease in QPS as the dataset size increases.
Part of the reason for this decrease is that a beam search with the
same parameters on a larger graph will not only be \textit{slower}
than the same search on a smaller graph, it will also be \textit{less
accurate} since it visits a much smaller fraction of all the vertices.
Since \cref{fig:qpsscale,fig:distcmpsscale} keep the recall fixed at
0.8, they must use an increased beam width at larger dataset sizes,
thus contributing to lower QPS.

\parhcnng{} and \parpynn{} both show steeper drops in QPS at fixed
recall than \pardiskann{} and \parhnsw{}.
This may be because they only express close neighbor relationships
with their edges.
As the data size grows, the relationships they express cover
smaller and smaller proportions of the whole dataset.
Thus, they require larger (more costly) parameters to obtain the same
level of recall as the data size increases.

Somewhat surprisingly, QPS and distance computations for FAISS
remained almost the same for the 100M and 1B datasets. We confirmed
that this phenomenon persisted through a wide range of parameter
choices.

In general, the non-graph based algorithms perform more distance computations but achieve
lower recall (and QPS).
This indicates that most of their distance computations are less effective than those in graph-based algorithms (i.e.,
were not contributing to finding closer neighbors).
This is possibly an important reason that they achieve much lower QPS than graph-based algorithms on a fixed recall,
and indicates the effectiveness of graph-based algorithms for ANNS.
\hide{
\laxman{seems to scale OK, but it's performing much worse than
the others}
\laxman{the bucketing techniques don't seem to do as well as the graph
based techniques since they waste a lot of distance comparisons by
looking at a whole bucket (many points within a bucket are
irrelevant)}

\laxman{TODO: add a summary about what we learned}
}


\subsection{Conclusions from Experiments}
\label{sec:exp:conclusion}
We summarize our findings about ANNS algorithms on billion scale
pointset below.

\begin{enumerate}[label=\arabic*.,topsep=0pt,itemsep=0pt,parsep=0pt,leftmargin=15pt]
	\item 
    Graph-based algorithms are especially capable at achieving high
    recall (greater than .9) at the scale of billions of points for
    QPS in the 10k--200k range.

    \item \faiss{} can achieve QPS close to the graph-based algorithms
    at a low recall, but QPS may significantly drop when a recall
    higher than 0.6 is required.

    \item The IVF algorithm \faiss{} struggled to achieve high recall at a billion scale, while \falconn{} achieved such low QPS that we did not include it in our experiments.


  \item All algorithms struggle to achieve high QPS on
  OOD data, but graph-based algorithms adapt much
  better: they can achieve 0.8 or higher recall with slightly
  lower QPS, while it is hard to achieve even 0.2 recall for IVF
  algorithms.

\end{enumerate}

\section{Related Work}\label{subsec: relatedwork}

\myparagraph{Approximate Nearest Neighbor Search Algorithms}
Data structures for ANNS fall roughly into four categories: graphs,
inverted indices, locality-sensitive hash tables, and trees. A
graph-based algorithm constructs a graph where the nodes represent
points in the index and the edges represent proximity relationships,
and where nearest neighbor queries are answered by applying a
heuristic search on the graph.
Prominent examples of graph-based algorithms include
NSG~\cite{fu2019nsg}, HNSW~\cite{malkov2020hnsw},
DiskANN~\cite{subramanya2019diskann}, but the academic literature
includes many other graph-based
approaches~\cite{munoz2019hcnng, fu2021high, zhang2022hierarchical,
	lu2021hvs, harwood2016fanng,
	dong2011efficient,mcinnes2020pynndescent,iwasaki2016pruned,
	iwasaki2018optimization,boytsov2013engineering,ann2016kgraph,chen2018sptag,ren2020hmann,opensearch2022opensearch,n22021n2,vespa2022vespa,kato2022vald}.

A commonly-used type of bucketing-based algorithms is the 
Inverted File Indexing (IVF) algorithms. 
IVF algorithms truncate the search space of a nearest neighbor
algorithm by partitioning vectors into buckets called \emph{posting
lists}; queries exhaustively search elements in only a small number of
lists instead of the entire space.  One assignment method is to use
a locality-sensitive hash (LSH) function. 
Inverted file
structures typically use a clustering algorithm to assign vectors to
posting lists, with distance to a representative element used to
determine which lists a query is mapped to. Some notable IVF-based
algorithms include PLSH~\cite{sundaram2013streaming},
FAISS-IVF~\cite{douze2011product, douze2016polysemous,
johnson2021billion}, and FALCONN~\cite{andoni2015practical},
along
with many others~\cite{gupta2022bliss, xia2013joint, pham2022falconn,
chen2021spann, aumuller2019puffin, klibisz2021tour,
opensearch2022opensearch, n22021n2}.

Trees such as $k$d-trees or cover trees are well-known data structures
for computing nearest neighbors in metric space with low
dimensionality (either actual or intrinsic)~\cite{beygelzimer2006cover,
arya1993fixed,gu2022parallel, klibisz2021tour}, useful for many such
applications~\cite{dobson2022parallel, connor2008parallel,
	yesantharao2021parallel}.
Their search methods are subject to the
curse of dimensionality,
\hide{\yihan{I don't think this applies to cover
trees? But it may not be as practical as others.}\laxman{can't they
still suffer if the intrinsic dimensionality is very large?}
}
but there are some modified tree-based
approaches for high dimensional search~\cite{muja2009fast,
	kula2019rpforest, scypy2022spatial, annoy2022approximate}.

In this paper, we focus on improving the scalability of building ANNS indexes based on graphs. 
There also exists work focusing on improving parallelism and scalability for other ANNS-related topics, such as
intra-query parallelism~\cite{peng2022speed,peng2023iqan} for graph-based algorithms, 
and improving scalability for tree-based algorithms on time series data~\cite{raoofy2023overcoming}.

\myparagraph{ANNS at a Billion Scale}
Next, we review what is currently known about scaling ANN algorithms to
billion-scale datasets.
Early work on ANN measured performance on datasets with up to a
billion points using various forms of
IVF~\cite{jegou2011searching,sundaram2013streaming,xia2013joint,baranchuk2018revisiting}.
The results for FAISS~\cite{douze2016polysemous}, the best known of the
algorithms in this class, have been reported for the \BIGANN{} and DEEP
billion scale datasets~\cite{faisswiki}.
These works do not include comparisons to graph-based algorithms, and
focus on recall for the single nearest neighbor instead of the $k$
nearest neighbors (i.e., 1@$n$ instead of $k$@$k$).

Other works use secondary storage-based algorithms to scale to
billion-scale datasets.  DiskANN~\cite{subramanya2019diskann}, a
graph-based algorithm, gives numbers for \BIGANN{} and DEEP
for a billion points.  They present limited comparisons to the
FAISS~\cite{douze2016polysemous} and IVFOADC+G+P
algorithms~\cite{baranchuk2018revisiting}.
The SPANN
system~\cite{chen2021spann} uses an inverted index where the posting
lists are stored in secondary memory.  On billion scale data (\BIGANN,
DEEP and \SPACEV) it only compares to DiskANN.
%
These existing works report the latency for one query at a time,
presumably because running multiple queries across cores does not
scale well due to limited secondary memory bandwidth and/or internal
parallelism within the query~\cite{chen2021spann}.
The query throughput is therefore much lower than in-memory-based
systems we report on in this paper, even accounting for machine size
(i.e., number of cores),  although they have the advantage of needing
less primary memory.

Johnson, Douze, and J\'{e}gou~\cite{johnson2021billion} report billion scale numbers on a
GPU-based implementation using an inverted-index-based approach.
Here again, the recall rates are low and the implementation is
only compared to another GPU-based system~\cite{Wieschollek16}.
Recent work on BLISS~\cite{gupta2022bliss} uses the same datasets as
we do at a billion scale.  They compare their approach to HNSW, but
the numbers they report for HNSW are much worse than those we
have found and that are reported here (over an order of magnitude).
Several systems work on a billion or more points, but do not report
numbers or comparisons to other
systems~\cite{fu2019nsg,malkov2020hnsw,opensearch2022opensearch,klibisz2021tour}.

\myparagraph{Benchmarking ANNS} There are two main works that
benchmark ANNS algorithms, one at the scale of millions of points and
one at the scale of billions. The first is the ANN Benchmarks
repository focusing on million-scale datasets~\cite{amuller2020ann}.
This is a benchmark suite of ANNS algorithms where any contributor may
submit an ANNS algorithm to be included in their public evaluations.
Each algorithm is run by the authors on up to nine million-scale
datasets.
Lastly, the Billion Scale ANNS Challenge, a competition hosted at
NeurIPS 2021~\cite{simhadri2021results}, focused on billion-scale ANNS
algorithms on three different hardware tracks
and six different billion-size datasets, including one range query dataset and
two datasets that exhibit OOD characteristics.
These existing benchmarks are a valuable resource, but their
user-sourced code for each algorithm is subject to implementation
differences and is not necessarily a comparison of the algorithmic
ideas.

\section{Conclusion and Future Work}\label{sec:conclusion}

We presented \sysname{}, which implements four parallel 
deterministic graph-based ANNS algorithms that scale to
billion-scale inputs on a single machine with high recall.
Our implementations avoid the use of locks, achieve better scalability
than existing implementations, and also outperformed existing
non-graph implementations in the ability of achieving high recall,
especially on OOD queries.
%

Our experiments illuminate many opportunities for future work. Here we
highlight some of the most interesting. One of our most surprising
conclusions is the strong performance of \hcnng{}, a relatively
lesser-known ANNS algorithm that does not appear in ANN Benchmarks.
This brings us to our first open question:

\begin{enumerate}[label=\textbf{Open Question \arabic*.},topsep=0pt,itemsep=0pt,parsep=0pt,listparindent=0pt, wide,series=open]
  \item Can the techniques from incremental graph algorithms be combined with insights from \hcnng{} to produce an algorithm which dominates both?
\end{enumerate}

\hide{
\medskip
Another of our most interesting conclusions is the clear superiority of IVF methods for range search, or, worded differently, the shocking inability of graph-based algorithms to adapt for range search. Due to ANNS graphs' ability to save distance comparisons by navigating quickly through the graph, our next two open questions ask how these insights can be used for range search:

\begin{enumerate}[resume*=open]
  \item How can ANNS graph structures be adapted to quickly answer range search queries?
\end{enumerate}

\medskip}

Another surprising result was the clear inability of IVF and LSH algorithms to answer out-of-distribution queries. This brings us to the next open problem:

\begin{enumerate}[resume*=open]
\item How can IVF and LSH algorithms be adapted to perform better on out-of-distribution queries?
\end{enumerate}

While our work focuses on comparison of indexing methods, quantization and/or compression of vector data is an important tool in approximate nearest neighbor search. Another open direction is:

\begin{enumerate}[resume*=open]
	\item How can quantization methods be efficiently parallelized and made deterministic, and how do such methods affect the choice of ANNS algorithms?
\end{enumerate}

Some closely-related problems to ANNS are \emph{range searches} (e.g., axis-align or fixed-radius, counting or reporting all, etc.). This brings us to the final open question:

\begin{enumerate}[resume*=open]
\item How do graph-based and other existing ANNS algorithms adapt to various range search problems at billion or larger scale?
\end{enumerate}

\begin{acks}                            
We thank the anonymous reviewers for their useful comments. Our experimental work was supported in part by Azure cloud compute credits granted by Microsoft Research. The authors were supported by NSF grants DGE1745016, DGE2140739, CCF-2103483, CCF-2238358, CCF-2227669, CCF-2119352, CCF-1919223 and CNS-2317194.
\end{acks}

\bibliographystyle{ACM-Reference-Format}
\bibliography{ann}

\appendix
\section{Algorithm Parameters}

\begin{figure*}[h]
	\begin{center}
		\small
		\begin{tabular}{c|c|c|c|c}
			\multicolumn{1}{c}{}& \multicolumn{1}{c}{\BIGANN} & \multicolumn{1}{c}{\SPACEV} & \multicolumn{1}{c}{\TTI} & \multicolumn{1}{c}{\SSNPP} \\
			\hline
			DiskANN & $R=64, L=128, \alpha=1.2$ & $R=64, L=128, \alpha=1.2$ & $R=64, L=128, \alpha=1.0$ & $R=150, L=400, \alpha=1.2$ \\
			\hline
			HNSW & $m=32, \mathit{efc}=128, \alpha=.82$ & $m=32, \mathit{efc}=128, \alpha=.83$ & $m=32, \mathit{efc}=128, \alpha=1.1$ & $m=75, \mathit{efc}=400, \alpha=.82$ \\
			\hline
			HCNNG & $T=30, Ls=1000, s=3$ & $T=50, Ls=1000, s=3$  & $T=30, Ls=1000, s=3$ & $T=50, Ls=1000, s=3$
			\\
			\hline
			pyNNDescent & \begin{tabular}{c} $K=40, Ls=100,$ \\$ T=10, \alpha=1.2$ \end{tabular}
			& \begin{tabular}{c} $K=60, Ls=100,$ \\$ T=10, \alpha=1.2$ \end{tabular} &   \begin{tabular}{c} $K=60, Ls=100,$ \\$ T=10, \alpha=.9$ \end{tabular} & \begin{tabular}{c} $K=60, Ls=1000,$ \\$ T=10, \alpha=1.4$ \end{tabular} \\
			\hline
			FAISS & \begin{tabular}{c} OPQ64\_128,\\IVF1048576\_HNSW32, \\PQ128x4fsr \end{tabular} & \begin{tabular}{c} OPQ64\_128,\\IVF1048576\_HNSW32,\\PQ64x4fsr \end{tabular} & \begin{tabular}{c}OPQ64\_128,\\IVF1048576\_HNSW32,\\PQ128x4fsr \end{tabular} & \begin{tabular}{c} OPQ64\_128,\\IVF1048576\_HNSW32,\\PQ64 \end{tabular} \\
			\hline
		\end{tabular}
	\end{center}
\caption{Parameters chosen for each dataset. For DiskANN, HCNNG, and pyNNDescent, $\alpha$ denotes the pruning parameter. For DiskANN, $R$ denotes degree bound and $L$ is the beam size. For HNSW, $m$ denotes the degree bound and $\mathit{efc}$ is the efConstruction. For HCNNG and pyNNDescent, $Ls$ denotes leaf size and $T$ denotes number of cluster trees. For HCNNG, $s$ denotes MST degree. For pyNNDescent, $K$ is the degree bound. For FAISS, the first string is the type of vector transform used for building the index, the second denotes the IVF index type, and the third indicates the PQ compression for the queries.}\label{fig: params}
\end{figure*}

\paragraph{\textbf{DiskANN}} The main parameters for the DiskANN index build are (1) the degree bound $R$, (2) the beam width $L$ used during insertion, and (3) the pruning parameter $\alpha$. In our experiments, we found that no single parameter setting was optimal for all recall regimes, and that there were significant tradeoffs in other recall values when maximizing for recall above .99; thus we chose to use parameters optimized for the .94-.97 range. Note that for \TTI, which minimizes negative inner product, the $\alpha$ value must be less than or equal to 1.0 in order to select for a denser graph.

\paragraph{\textbf{HNSW}} The parameters for the HNSW index build have similar meanings to those of DiskANN where (1) $m$ is the general degree bound, (2) $\mathit{efc}$ is the beam width for index build, and (3) $\alpha$ controls the graph density. Specifically, the bottom layer has degree bound $2m$ while all the other layers have degree bound $m$. This design is referred to in the source code of hnswlib~\cite{hnswlib2019hnswlib} and performs better in practice than setting all layers to have the same degree bound. To make the results of DiskANN and HNSW comparable, we keep $2m=R$ and $\mathit{efc}=L$ through all the datasets and adjust $\alpha$ to reach similar average degrees at the bottom layer.

\paragraph{\textbf{HCNNG}} The relevant parameters for the HCNNG index build are (1) the leaf size $Ls$ of the random clustering tree, (2) the maximum degree $s$ of the MST built in each leaf, and (3) the number $T$ of random clustering trees. In our experiments we found that a leaf size of $1000$ sufficed for all dataset sizes. We use the original authors' suggested parameter of $3$ for the maximum degree of the MST. In our experiments we found that $30-50$ trees sufficed for our datasets, and we found that QPS began to increase after more than $50$ trees.

\paragraph{\textbf{pyNNDescent}} The relevant parameters for the pyNNDescent index build are (1) the degree bound $K$, (2) the pruning parameter $\alpha$, (3) the number $T$ of random clustering trees used to seed the initial graph, and (4) the leaf size $Ls$ of the random clustering trees. In our experiments we found that a degree bound of $40-60$ worked for most datasets.

\begin{figure}
	\centering
	\includegraphics[width=\columnwidth]{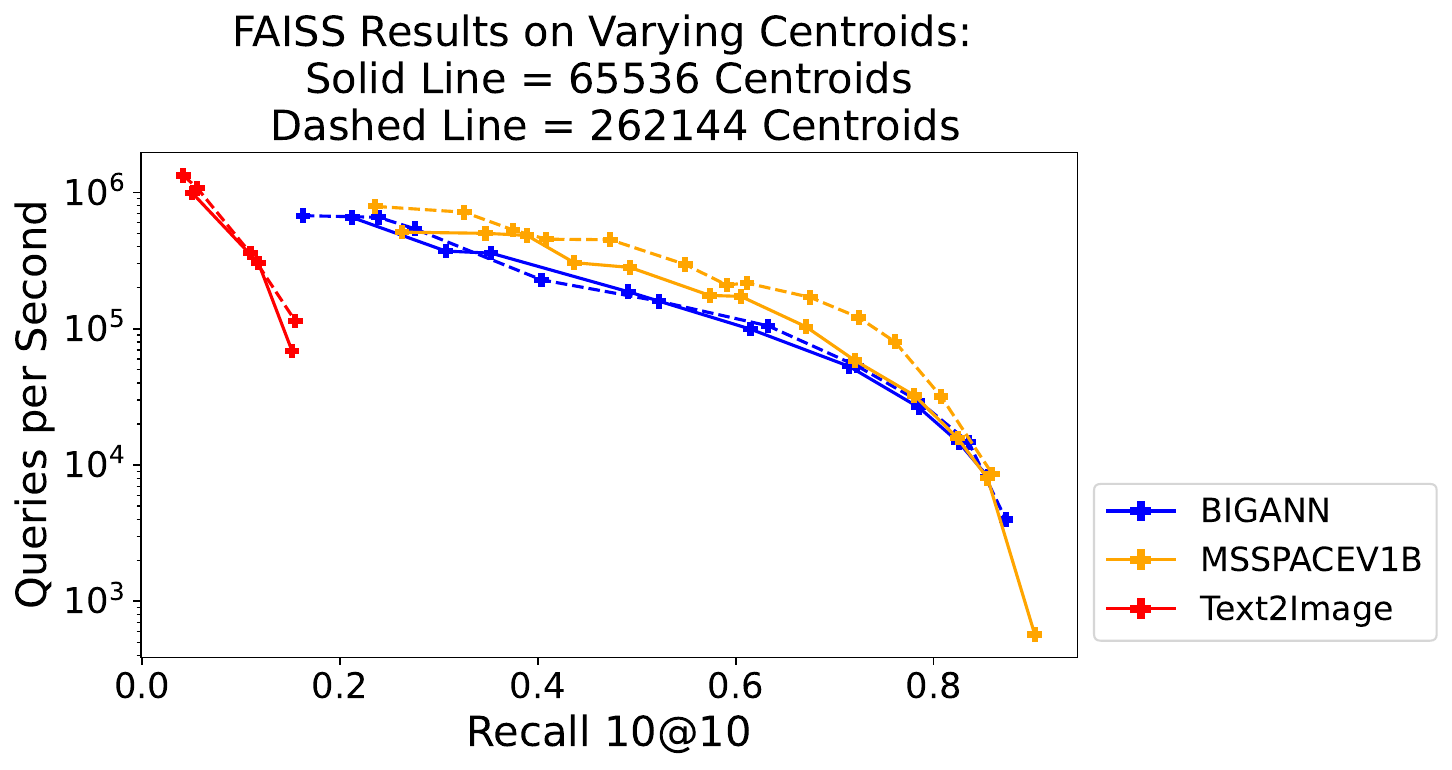}
	\vspace{-1em}
	\caption{ QPS on FAISS builds with varying centroids; solid lines indicate $2^{16}$ centroids and dashed lines indicate $2^{18}$ centroids. All builds are on a dataset size of 100 million and QPS is calculated using 96 threads. }\label{fig:FAISSCentroids}
\end{figure}

\paragraph{\textbf{FAISS}} The three main aspects of FAISS index construction are (1) the PQ vector transform for the index build, (2) the IVF index, and (3) the PQ compression for the queries. To choose index parameters for FAISS, we referred to the index parameters published in the Big ANN Benchmarks competition~\cite{simhadri2021results} as well as the optimal parameters for \BIGANN{} published on the FAISS Wiki~\cite{faisswiki}. Since these parameters optimized for small space usage (limiting the index size to 128 GB for a billion-scale build), we experimented with increasing the number of bits in the PQ compression as well as increasing the number of posting lists in order to optimize for the high-recall region. In Figure~\ref{fig:FAISSCentroids}, we show the effects of increasing the number of centroids on the 100M slices of all datasets. In Figure~\ref{fig: params} we show the parameters chosen for billion-scale builds. For hundred-million scale builds, there is no single dominant option for \SPACEV{} and \BIGANN{}, so we include two FAISS builds in Figure~\ref{fig:100Mresults}.

\end{document}


\title[Supplemental Material]{Supplemental Material}         

\maketitle

\section{Algorithm Descriptions}

\subsection{Inverted Indexing Algorithms}
Next, we introduce inverted-indexing algorithms (IVF), which
constitute a large class of popular algorithms based on
locality-sensitive hashing (LSH) and clustering. At a high level, IVF
algorithms partition the vectors into buckets called \emph{posting
	lists}; at query time the query only exhaustively searches elements in
only a small number of lists instead of the entire space. We study two
IVF-based algorithms in-depth in this paper, which we describe next.

\paragraph{\textbf{FAISS-IVF}} FAISS is a widely-used and highly optimized library for efficient similarity searching on CPUs and
GPUs~\cite{jegou2011searching, douze2011product, douze2016polysemous,
	johnson2021billion}. It provides a variety of options for
\textit{product quantization (PQ)}---that is, compressing vectors to a
lower dimensionality to make distance computations faster. The primary
search structure of FAISS is the inverted index; points are typically
assigned to buckets via single-level clustering or two-level
clustering, but can add other techniques such as using HNSW or NSG to
help select buckets during search. At the billion scale, a typical
FAISS index follows three steps: (1) compute a PQ transform on the
vectors for the purpose of building the index, (2) build an inverted
file index, possibly with an HNSW structure over the buckets, and (3)
compute a PQ compression of the vectors for the purposes of querying
the index.

We use the original implementation of FAISS, through the BigANN
Benchmarks testing framework~\cite{simhadri2021results}, in our
experiments.
\vspace{-.5 em}

\paragraph{\textbf{FALCONN}} FALCONN is a C++ library for LSH-based ANNS
algorithms~\cite{andoni2015practical}.
Locality-sensitive hash (LSH) functions are class of hash functions that can easily cause collisions for close vectors in a space.
More specifically, such hash functions perform randomized space partitions of a unit sphere with the same dimension as the input, and mark the vectors in the same partition (as a bucket) with the identical hash value.
FALCONN uses multiple hash functions to create each hash table, thus increasing the number of buckets and decreasing bucket size. This prevents buckets from having too many elements and thus prevents the search from having to access too many candidates. To improve accuracy, it builds multiple (replicated) hash tables for higher probability of success, trading off accuracy for higher memory consumption. A search hashes the queried point using each hash function, then collects all the elements from corresponding buckets among the tables, and retains the nearest neighbors among these candidates.
Furthermore, by enabling multi-probe LSH, which assigns a vector to more than one bucket in each hash table,
FALCONN is able to go consider more candidates from the additional buckets without needing to create more hash tables, thus both improving accuracy and saving memory.

\subsection{Algorithms Excluded}\label{sec:excluded} Out of the
algorithms appearing in ANN-benchmarks~\cite{amuller2020ann} and cited
in this work, we excluded algorithms for one of two reasons: either
the algorithm was consistently in the bottom half of the QPS
distribution for most datasets, or we already included an algorithm
with a similar approach.  Algorithms excluded on the basis of
similarity to DiskANN include NSG~\cite{fu2019nsg},
NSSG~\cite{fu2021high}, FANNG~\cite{harwood2016fanng},
PANNG~\cite{iwasaki2016pruned}, ONNG~\cite{iwasaki2018optimization},
Vald~\cite{kato2022vald}, and kGraph~\cite{ann2016kgraph}, as they are all incremental greedy search-based algorithms producing a flat graph.
Algorithms excluded on the basis of similarity to HNSW (including
re-implementations of HNSW appearing in
ANN-benchmarks) include HSSG~\cite{zhang2022hierarchical},
HVS~\cite{lu2021hvs}, Proximity Graph
Maintenance~\cite{xu2022proximity}, HM-ANN~\cite{ren2020hmann},
n2~\cite{n22021n2}, and Vespa~\cite{vespa2022vespa}, as they are incremental algorithms which produce a layered graph. Algorithms
excluded on the basis of similarity to FAISS-IVF include
BLISS~\cite{gupta2022bliss}, Joint Inverted
Indexing~\cite{xia2013joint}, SPANN~\cite{chen2021spann}, and
SCaNN~\cite{guo2020accelerating} as they are other IVF methods. Algorithms excluded on the basis of
similarity to FALCONN include PLSH~\cite{sundaram2013streaming},
FALCONN++~\cite{pham2022falconn}, PUFFIN~\cite{aumuller2019puffin},
and ElastiKNN~\cite{klibisz2021tour} as they are other LSH methods. The
NNDescent~\cite{dong2011efficient} algorithm was excluded on basis of
similarity to pyNNDescent. The SPTAG~\cite{chen2018sptag} algorithm
was excluded on basis of similarity to HCNNG. Algorithms excluded due
to low performance in the benchmark suite include
FLANN~\cite{muja2009fast}, RPForest~\cite{kula2019rpforest},
ANNOY~\cite{annoy2022approximate}, OpenSearch
KNN~\cite{opensearch2022opensearch}, and ScyPy's
ckdTree~\cite{scypy2022spatial}.

\section{Algorithm Parameters}

\begin{figure*}
	\begin{center}
		\small
		\begin{tabular}{c|c|c|c|c}
			\multicolumn{1}{c}{}& \multicolumn{1}{c}{\BIGANN} & \multicolumn{1}{c}{\SPACEV} & \multicolumn{1}{c}{\TTI} & \multicolumn{1}{c}{\SSNPP} \\
			\hline
			DiskANN & $R=64, L=128, \alpha=1.2$ & $R=64, L=128, \alpha=1.2$ & $R=64, L=128, \alpha=.98$ & $R=150, L=400, \alpha=1.2$ \\
			\hline
			HNSW & $m=32, \mathit{efc}=128, \alpha=.82$ & $m=32, \mathit{efc}=128, \alpha=.83$ & $m=32, \mathit{efc}=128, \alpha=1.1$ & $m=75, \mathit{efc}=400, \alpha=.82$ \\
			\hline
			HCNNG & $T=30, Ls=1000, s=3$ & $T=50, Ls=1000, s=3$  & $T=30, Ls=1000, s=3$ & $T=50, Ls=1000, s=3$
			\\
			\hline
			pyNNDescent & \begin{tabular}{c} $K=40, Ls=100,$ \\$ T=10, \alpha=1.2$ \end{tabular}
			& \begin{tabular}{c} $K=60, Ls=100,$ \\$ T=10, \alpha=1.2$ \end{tabular} &   \begin{tabular}{c} $K=60, Ls=100,$ \\$ T=10, \alpha=.9$ \end{tabular} & \begin{tabular}{c} $K=60, Ls=1000,$ \\$ T=10, \alpha=1.4$ \end{tabular} \\
			\hline
			FAISS & \begin{tabular}{c} OPQ64\_128,\\IVF1048576\_HNSW32, \\PQ128x4fsr \end{tabular} & \begin{tabular}{c} OPQ64\_128,\\IVF1048576\_HNSW32,\\PQ64x4fsr \end{tabular} & \begin{tabular}{c}OPQ64\_128,\\IVF1048576\_HNSW32,\\PQ128x4fsr \end{tabular} & \begin{tabular}{c} OPQ64\_128,\\IVF1048576\_HNSW32,\\PQ64 \end{tabular} \\
			\hline
			FALCONN &$l=30, rot=1$ &$l=30, rot=1$ &$l=30, rot=1$ &$l=30, rot=1$ \\
			\hline
		\end{tabular}
	\end{center}
	\vspace{-1em}
	\caption{ Parameters chosen for each dataset. \normalfont{For DiskANN, HCNNG, and pyNNDescent, $\alpha$ denotes the pruning parameter. For DiskANN, $R$ denotes degree bound and $L$ is the beam size. For HNSW, $m$ denotes the degree bound and $\mathit{efc}$ is the efConstruction. For HCNNG and pyNNDescent, $Ls$ denotes leaf size and $T$ denotes number of cluster trees. For HCNNG, $s$ denotes MST degree. For pyNNDescent, $K$ is the degree bound. For FAISS, the first string is the type of vector transform used for building the index, the second denotes the IVF index type, and the third indicates the PQ compression for the queries. For FALCONN, $l$ is the number of hash tables, and $\mathit{rot}$ is the number of rotations.}} \label{fig:params}
\end{figure*}

\paragraph{\textbf{DiskANN}} The main parameters for the DiskANN index build are (1) the degree bound $R$, (2) the beam width $L$ used during insertion, and (3) the pruning parameter $\alpha$. In our experiments, we found that no single parameter setting was optimal for all recall regimes, and that there were significant tradeoffs in other recall values when maximizing for recall above .99; thus we chose to use parameters optimized for the .94-.97 range. Note that for \TTI, which minimizes negative inner product, the $\alpha$ value must be less than one in order to select for a denser graph.


\paragraph{\textbf{HNSW}} The parameters for the HNSW index build have similar meanings to those of DiskANN where (1) $m$ is the general degree bound, (2) $\mathit{efc}$ is the beam width for index build, and (3) $\alpha$ controls the graph density. Specifically, the bottom layer has degree bound $2m$ while all the other layers have degree bound $m$. This design is referred to in the source code of hnswlib~\cite{hnswlib2019hnswlib} and performs better in practice than setting all layers to have the same degree bound. To make the results of DiskANN and HNSW comparable, we keep $2m=R$ and $\mathit{efc}=L$ through all the datasets and adjust $\alpha$ to reach similar average degrees at the bottom layer.

\paragraph{\textbf{HCNNG}} The relevant parameters for the HCNNG index build are (1) the leaf size $Ls$ of the random clustering tree, (2) the maximum degree $s$ of the MST built in each leaf, and (3) the number $T$ of random clustering trees. In our experiments we found that a leaf size of $1000$ sufficed for all dataset sizes. We use the original authors' suggested parameter of $3$ for the maximum degree of the MST. In our experiments we found that $30-50$ trees sufficed for our datasets, and we found that QPS began to increase after more than $50$ trees.

\paragraph{\textbf{pyNNDescent}} The relevant parameters for the pyNNDescent index build are (1) the degree bound $K$, (2) the pruning parameter $\alpha$, (3) the number $T$ of random clustering trees used to seed the initial graph, and (4) the leaf size $Ls$ of the random clustering trees. In our experiments we found that a degree bound of $40-60$ worked for most datasets.

\begin{figure}
	\centering
	\includegraphics[width=\columnwidth]{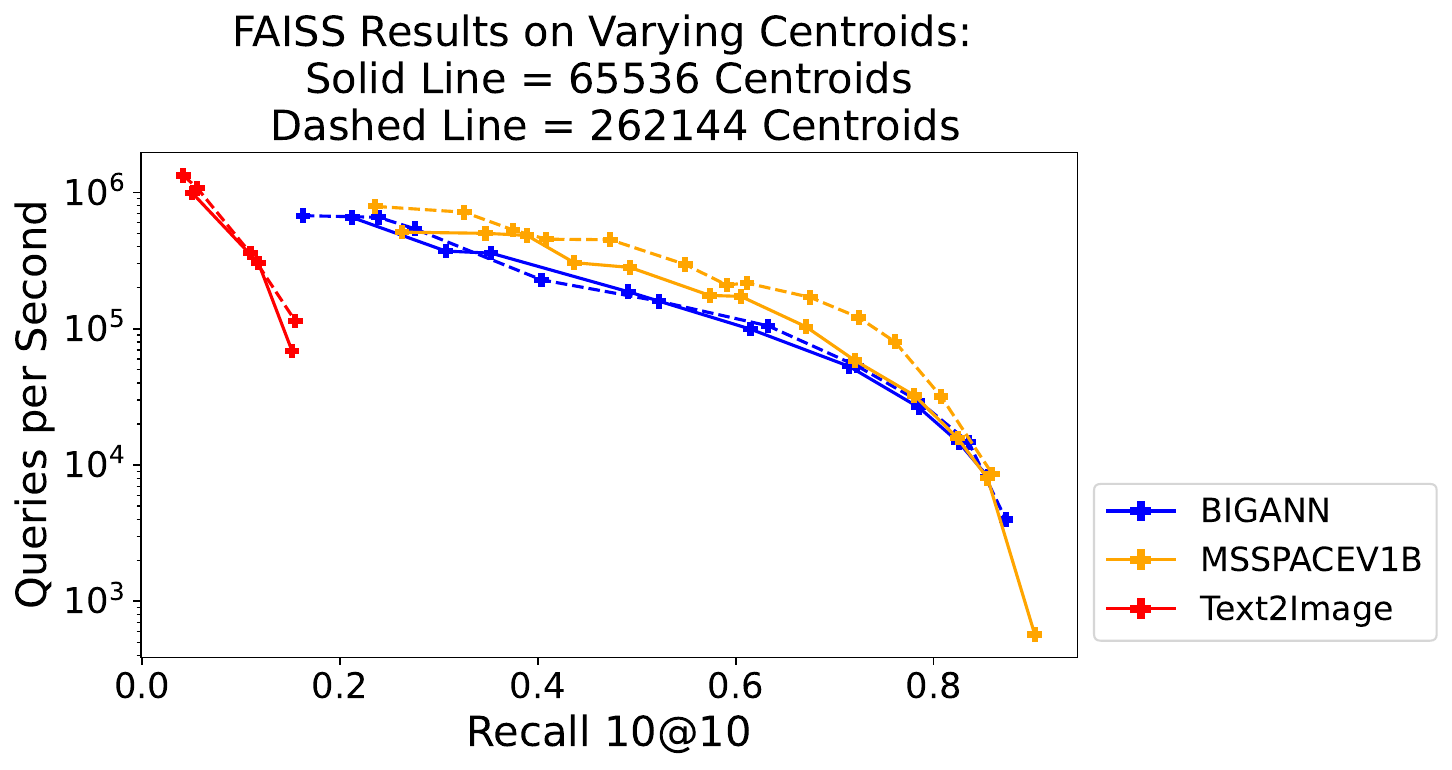}
	\vspace{-1em}
	\caption{ QPS on FAISS builds with varying centroids; solid lines indicate $2^{16}$ centroids and dashed lines indicate $2^{18}$ centroids. All builds are on a dataset size of 100 million on the \evf machine. }\label{fig:FAISSCentroids}
	
\end{figure}

\begin{figure*}[t]
	\begin{subfigure}[b]{.31\textwidth}
		\centering
		\includegraphics[width=\textwidth]{matplotlib/QPSBIGANN100M_ev5}
		
		\caption{\BIGANN-100M}\label{fig:bigannQPSEv5}
	\end{subfigure}
	\hfill
	\begin{subfigure}[b]{.31\textwidth}
		\centering
		\includegraphics[width=\textwidth]{matplotlib/QPSbigannHigh}
		
		\caption{\BIGANN-100M in high recall regime}\label{fig:bigannQPSEv5High}
	\end{subfigure}
	\hfill
	\begin{subfigure}[b]{.31\textwidth}
		\centering
		\includegraphics[width=\textwidth]{matplotlib/QPST2I100M_ev5}
		
		\caption{\TTI-100M}\label{fig:text2imageQPSEv5}
	\end{subfigure}
	\begin{subfigure}[b]{.31\textwidth}
		\centering
		\includegraphics[width=\textwidth]{matplotlib/QPSSPACEV100M_ev5}
		
		\caption{\SPACEV-100M}\label{fig:spacevQPSEv5}
	\end{subfigure}
	\hfill
	\begin{subfigure}[b]{.31\textwidth}
		\centering
		\includegraphics[width=\textwidth]{matplotlib/qpsspacevhigh}
		
		\caption{\SPACEV-100M in high recall regime}\label{fig:spacevQPSEv5High}
	\end{subfigure}
	\hfill
	\begin{subfigure}[b]{.31\textwidth}
		\centering
		\includegraphics[width=\textwidth]{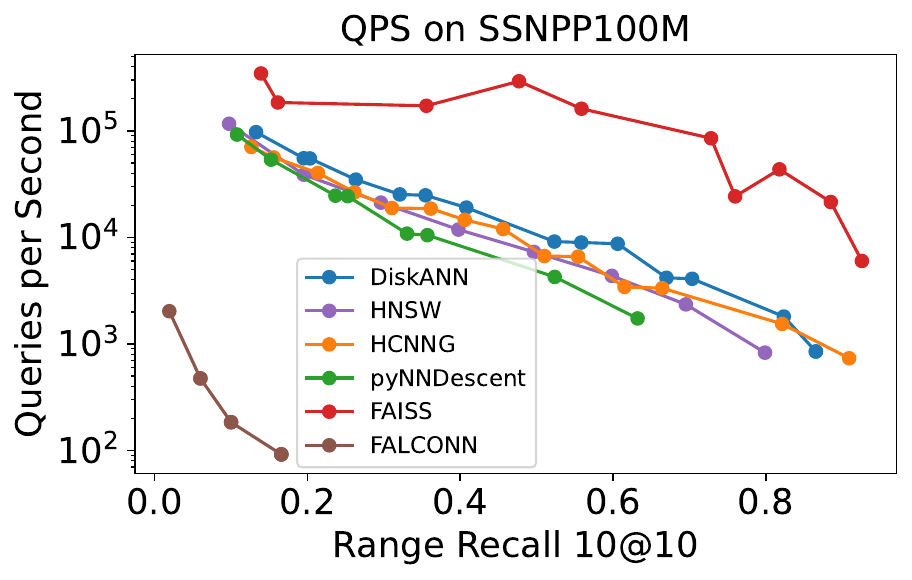}
		
		\caption{\SSNPP-100M, Range Search.}\label{fig:ssnppQPSEv5}
	\end{subfigure}
	\caption{QPS-recall curves on all 100-million size datasets on the \evf{} machine (QPS as a function of recall).}\label{fig:100Mresults}
\end{figure*}

\paragraph{\textbf{FAISS}} The three main aspects of FAISS index construction are (1) the PQ vector transform for the index build, (2) the IVF index, and (3) the PQ compression for the queries. To choose index parameters for FAISS, we referred to the index parameters published in the Big ANN Benchmarks competition~\cite{simhadri2021results} as well as the optimal parameters for \BIGANN{} published on the FAISS Wiki~\cite{faisswiki}. Since these parameters optimized for small space usage (limiting the index size to 128 GB for a billion-scale build), we experimented with increasing the number of bits in the PQ compression as well as increasing the number of posting lists in order to optimize for the high-recall region. In Figure~\ref{fig:FAISSCentroids}, we show the effects of increasing the number of centroids on the 100M slices of all datasets. In Figure~\ref{fig:params} we show the parameters chosen for billion-scale builds. For hundred-million scale builds, there is no single dominant option for \SPACEV{} and \BIGANN{}, so we include two FAISS builds in Figure~\ref{fig:100Mresults}.
\paragraph{\textbf{FALCONN}} We sweep the index parameters in reasonable ranges using the documentation in the FALCONN code repository \cite{falconn2017falconn} and choose the best combination in our observation. There are many parameters for FALCONN to build hash tables. Considering that we focus on large datasets, the hash family is fixed to cross-polytope LSH for better performance on large datasets. The number of hash tables, $l$, is a core parameter that affects the building time, query accuracy, and memory usage. In our experiments, we found that $l=30$ maximizes accuracy within the available memory capacity. The number of hash functions per table follows the recommended setting from the helper functions in the authors' code, computed based on the input size and the dimension. We also tried manually setting it to other values but they did not perform better.

\section{Additional Experimental Results}

\begin{figure}
	\centering
	\includegraphics[width=.8\columnwidth]{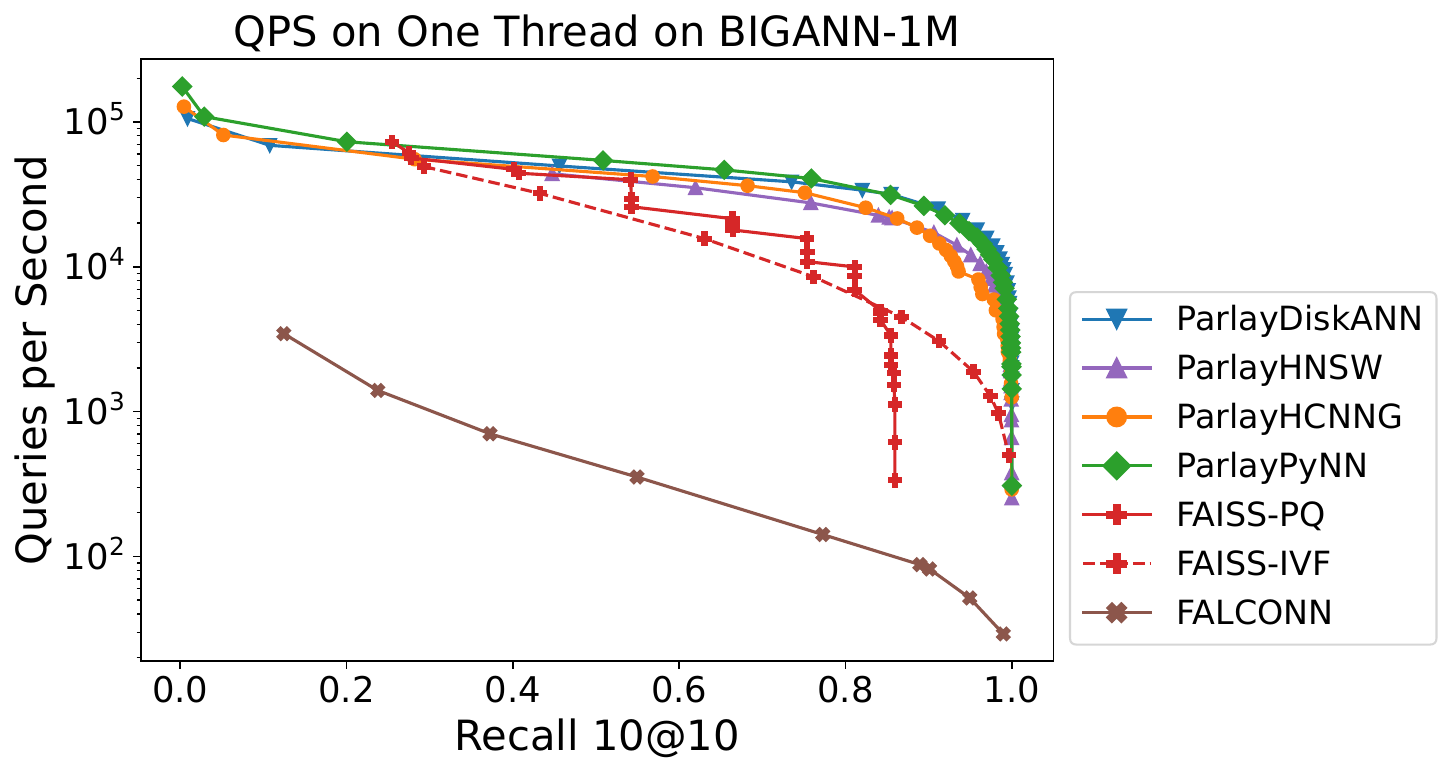}
	\vspace{-1.5em}
	\caption{ QPS on a single thread on \BIGANN-1M on \evf. Shown to compare with ann-benchmarks.}
	\label{fig:baseline}	
\end{figure}

\begin{figure}[t]
	\centering
	\includegraphics[width=\columnwidth]{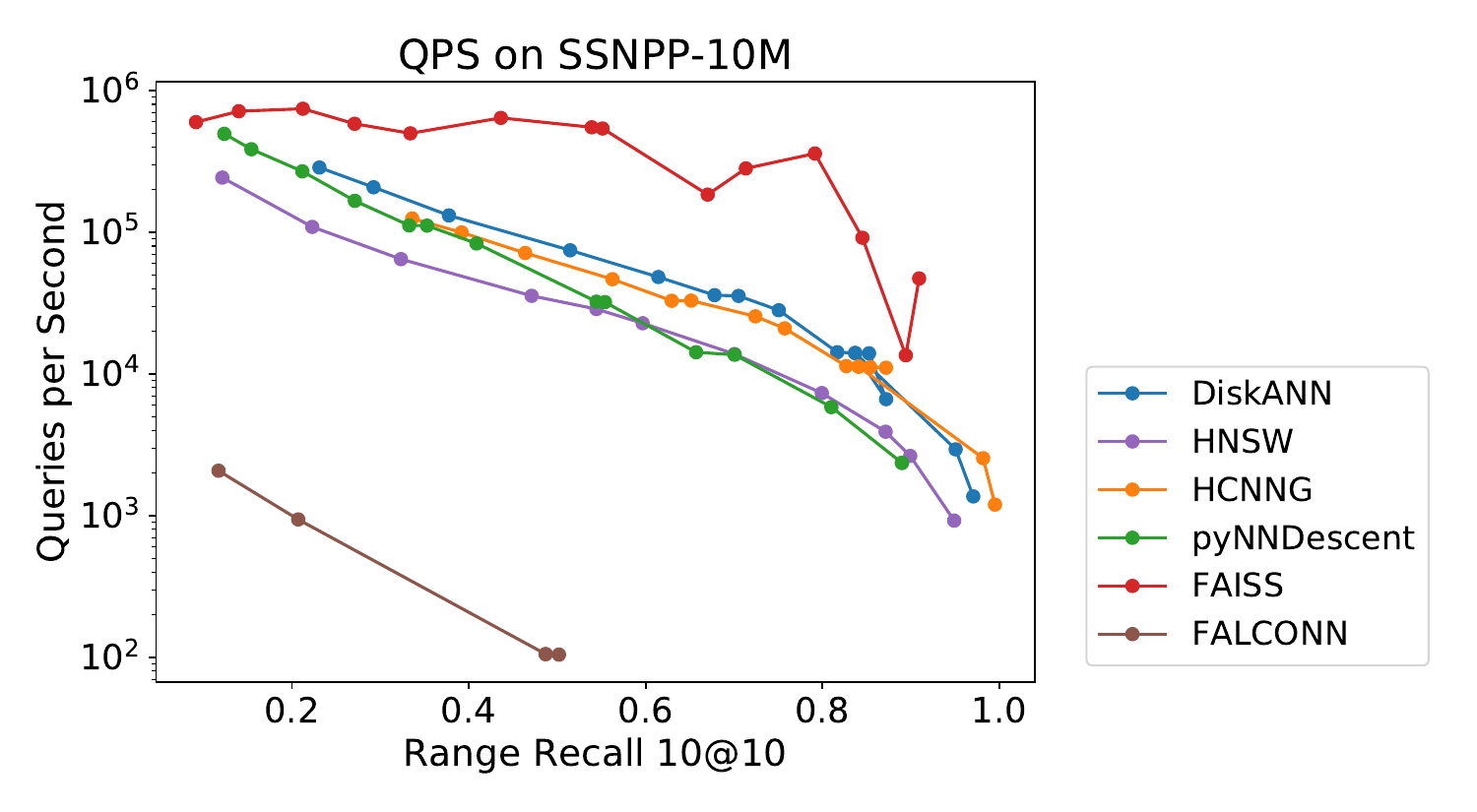}
	\vspace{-2em}
	\caption{ QPS as a function of recall on range search dataset \SSNPP-10M.}\label{fig:rangeqps10M}
\end{figure}

\begin{table}[t]
	\begin{center}
	\begin{tabular}{c@{  }@{  }c@{  }@{  }c@{  }@{  }c@{  }@{  }c}
		& \BIGANN & \SPACEV & \TTI & \SSNPP \\
		\hline
		DiskANN & .89 & 1.12 & 1.35 & 9.27 \\
		HNSW & .92 & .77 & 1.79 & 11.85 \\
		HCNNG & .74 & 1.22 & 2.43 & 1.52 \\
		pyNNDescent & .52 & .83 & 2.43 & 1.52 \\
		FAISS & .28 & .17 & .29 & .21 \\
		FALCONN & .06  & .06 & .08 & .08 \\
		\hline
	\end{tabular}
\end{center}
\caption{ Build times (hours) on hundred million scale datasets.}\label{fig:buildtimeshundredmil}
\end{table}

\begin{figure*}[t]
	
	\begin{subfigure}[b]{.24\textwidth}
		\centering
		\includegraphics[width=\textwidth]{matplotlib/DistCbigann}
		
		\caption{ BIGANN-1B}\label{fig:biganndistcmps}
	\end{subfigure}
	\hfil
	\begin{subfigure}[b]{.24\textwidth}
		\centering
		\includegraphics[width=\textwidth]{matplotlib/DistCspacev}
		
		\caption{ SPACEV-1B}\label{fig:spacevdistcmps}
	\end{subfigure}
	\hfil
	\begin{subfigure}[b]{.24\textwidth}
		\centering
		\includegraphics[width=\textwidth]{matplotlib/DistCmpsT2I}
		
		\caption{ TTI-1B}\label{fig:ooddistcmps}
	\end{subfigure}
	\hfil
	\begin{subfigure}[b]{.24\textwidth}
		\centering
		\includegraphics[width=\textwidth]{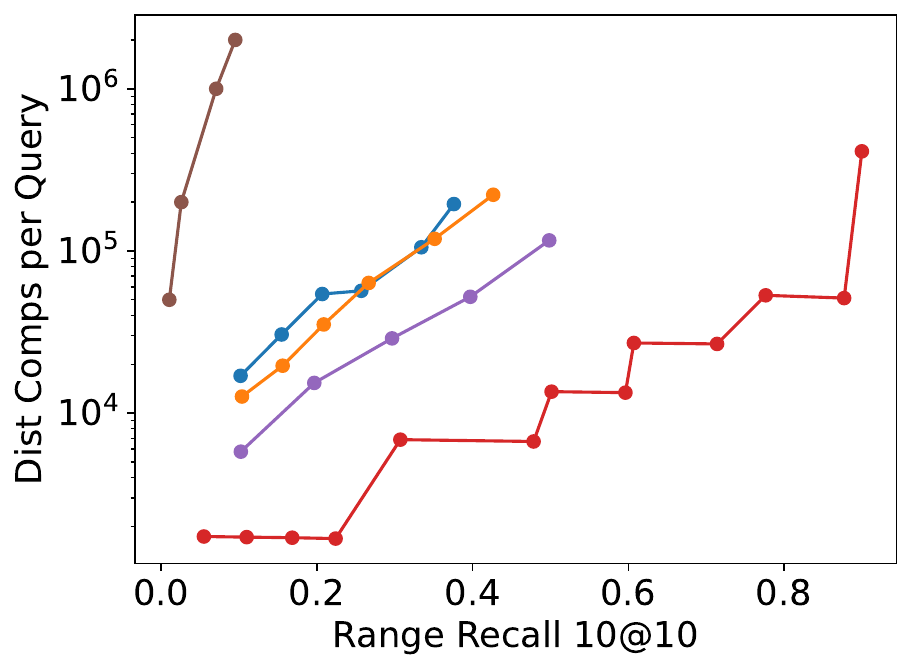}
		
		\caption{\SSNPP-1B, Range Search}\label{fig:rangedistcmps}
	\end{subfigure}
	
	\textbf{Distance comparisons per query as a function of recall.}
	
	\caption{ Graphs showing QPS in the high and low recall regimes as well as distance comparisons per query on all billion-size datasets on the \msvt{} machine. }
	\label{fig:billionsize}
	
\end{figure*}

\bibliography{ann}